\documentclass[11pt]{article}

\usepackage[margin=1in]{geometry}
\usepackage{amsmath,amssymb,amsthm,mathtools}
\usepackage{bbm}
\usepackage{natbib}
\usepackage{setspace}
\usepackage{microtype}
\usepackage{xcolor}
\newif\ifreviewchanges
\reviewchangestrue 

\definecolor{jglpurple}{RGB}{128,0,160}

\usepackage[textsize=scriptsize]{todonotes} 

\usepackage{authblk}
\usepackage{pgfplots}
\usetikzlibrary{arrows.meta,calc}
\pgfplotsset{compat=1.18}
\usepgfplotslibrary{groupplots}
\pgfplotsset{paperplot/.style={scale only axis,axis lines*=left,tick align=outside,typeset ticklabels with strut,tick label style={font=\footnotesize},label style={font=\small},clip=false}}
\usepackage{hyperref}

\newcommand{\R}{\mathbb{R}}
\newcommand{\E}{\mathbb{E}}
\renewcommand{\P}{\mathbb{P}}
\newcommand{\cL}{\mathcal{L}}
\newcommand{\cF}{\mathcal{F}}
\newcommand{\ES}{\mathrm{ES}}
\newcommand{\VaR}{\mathrm{VaR}}
\newcommand{\Var}{\mathrm{Var}}
\newcommand{\eqd}{\stackrel{\smash{\mathrm{d}}}{=}}
\newcommand{\lecx}{\le_{\mathrm{cx}}}
\DeclareMathOperator*{\esssup}{ess\,sup}

\usepackage{xcolor}

\theoremstyle{plain}
\newtheorem{theorem}{Theorem}
\newtheorem{proposition}{Proposition}
\newtheorem{corollary}{Corollary}
\newtheorem{lemma}{Lemma}
\theoremstyle{definition}
\newtheorem{assumption}{Assumption}
\newtheorem{definition}{Definition}
\newtheorem{example}{Example}

\newtheoremstyle{axiomstyle}{\topsep}{\topsep}{\itshape}{}{\bfseries}{.}{.5em}{\thmname{#1}\thmnote{ #3}}
\theoremstyle{axiomstyle}
\newtheorem*{axiom}{Axiom}
\newtheorem*{property}{Property}

\title{An axiomatic foundation for optimized certainty equivalents}
\author[1]{Christopher Blier-Wong}
\author[2]{Jean-Gabriel Lauzier}
\author[3]{Zachary Van Oosten}

\affil[1]{Department of Statistical Sciences, University of Toronto, Canada}
\affil[2]{Department of Economics, Memorial University of Newfoundland, Canada}
\affil[3]{Department of Decision Sciences, HEC Montr\'eal, Canada}

\date{\today}

\begin{document}

\maketitle

\begin{abstract}
\noindent Optimized Certainty Equivalents (OCEs) are an important class of risk measures for optimization and risk management that includes the Mean--Variance criterion, Expected Shortfall, the Entropic Risk Measure, and expectation. Despite their importance, economic axioms have not characterized OCEs. We propose a new axiom called Fair-Coin Compensation, which characterizes OCEs together with Strong Risk Aversion and Normalization. We interpret Fair-Coin Compensation through the fair assignment of two losses by a coin toss: transfers between two participants can remove the additional risk created by the random assignment. Adding further axioms yields sharper characterizations. Monotonicity yields standard OCEs; Positive Homogeneity, together with Prudence, yields Expected Shortfall; and Fair-Coin Independence yields the Entropic Risk Measure. We also axiomatize the Mean--Variance criterion using our novel axiom of Equal-Mean Randomization, which states that random assignment of two losses with the same expected value does not produce a risk measure above the average of their risk measures.

\end{abstract}
\section{Introduction}\label{sec:intro}

\citet{BenTal1986} introduced the optimized certainty equivalent (OCE) as the certainty equivalent resulting from optimally allocating uncertain income between present and future consumption. Let the random variable $X$ represent a loss, so that higher positive values are undesirable. The OCE of $X$ with loss function $\ell$ is
\begin{equation}\label{eq:oce}
\rho_\ell(X)=\inf_{m\in\R}\left\{m+\E[\ell(X-m)]\right\},
\end{equation}
where $m$ is the cash amount paid in advance, $X-m$ is the part of the loss this amount does not cover, and $\ell$ is a convex function that penalizes the residual loss. When $\ell$ is nondecreasing, \eqref{eq:oce} defines a standard OCE in the sense of \citet{BenTal1986}. \citet{BenTal2007} showed that standard OCEs are convex risk measures \citep{Follmer2002, Frittelli2002} and that the penalty in their dual representation is a $\varphi$-divergence between an alternative probability measure and the reference probability measure.

The class of OCEs is important for risk management and optimization. First, several important risk measures used in practice are OCEs, including the Mean--Variance criterion, Expected Shortfall, the Entropic Risk Measure, and expectation. For Expected Shortfall, \eqref{eq:oce} is the minimization formula of \citet{Rockafellar2000, Rockafellar2002}, in which Value-at-Risk is an optimal cash amount $m$. Second, \eqref{eq:oce} is convenient in optimization. To minimize $\rho_\ell$ over a set of decisions, one can jointly minimize over the decisions and the cash amount $m$; the joint objective is an expected value that can be estimated by averaging over scenarios and minimized using stochastic optimization methods. For Expected Shortfall, \citet{Rockafellar2000} use this property to reduce the minimization to a linear program. Standard OCEs appear as the convex Bayes risk measures of \citet{Embrechts2021} and as objectives in risk-sensitive learning \citep{Lee2020}. Their dual representation connects them to the $\varphi$-divergence penalties used in distributionally robust optimization \citep{BenTal2013} and to the divergence preferences of \citet{Maccheroni2006} in decision theory.

Given the class's central role in decision theory and risk management, it is natural to ask which axioms characterize it, that is, which principles lead an institution to choose its risk measure within the family of OCEs. This paper provides an axiomatic characterization of OCEs. Characterizing the class is challenging because the average of two OCEs need not be an OCE. Every OCE is cash additive, convex in outcome mixtures, law invariant, consistent with the convex order, and concave in probabilistic mixtures, and each of these properties is preserved under averaging. A characterization of OCEs therefore requires an axiom that averaging does not preserve. 

We introduce a new economic axiom called \textit{Fair-Coin Compensation}, and show in Theorem~\ref{thm:main} that it pins down OCEs together with the two standard axioms of Strong Risk Aversion \citep{Rothschild1970} and (cash) Normalization. Fair-Coin Compensation compares the risk of assigning two losses by a fair coin with the risks of the two losses on their own. Suppose that two losses $X$ and $Y$ must be assigned to two financial institutions that assess risk with the same risk measure, and that the assignment is made by tossing an independent fair coin, so each participant receives $X$ or $Y$ with probability one half. One loss may be more severe than the other, so each participant faces the risk of receiving the more severe loss. Suppose that the institutions can agree to a transfer before the coin is tossed to mitigate the surplus risk generated by this random assignment. That is, they agree on a side payment $a$ that depends only on the coin's outcome, where the institution that receives $X$ pays $a$ to the institution that receives $Y$. Fair-Coin Compensation requires that the minimum risk evaluation attainable through such transfers equals the average of the standalone risk evaluations of $X$ and $Y$. Formally, let $B=1$ if the coin assigns $X$ to the institution and $B=0$ otherwise. Then
\begin{equation}\label{eq:FCintro}
\inf_{a\in\R}\rho\left(B(X+a)+(1-B)(Y-a)\right)=\frac{\rho(X)+\rho(Y)}{2}.
\end{equation}

Random assignment is a long-established economic mechanism for allocating indivisible resources fairly, for example in school choice and course allocation \citep{Budish2013}, and for breaking ties between the highest bids in auctions \citep{myerson1981}. However, it need not be costless for participants who are averse to probabilistic mixtures \citep{Dillenberger2021, Dillenberger2025}. The equality in~\eqref{eq:FCintro} implies two properties of this cost. First, the random assignment does not reduce the risk measure below the average of the standalone risk measures. This is an aversion to random assignment called concavity in probabilistic mixtures that many risk measures used in practice satisfy. Second, a side payment can make the additional risk from the random assignment arbitrarily small. This second requirement singles out the OCEs among risk measures. For example, a single Expected Shortfall satisfies this property, but an average of two Expected Shortfalls at different levels fails it and is thus not an OCE.

The three axioms characterizing OCEs are independent: none is implied by the other two. In Proposition~\ref{prop:properties}, we show that they imply the risk measure is law invariant, cash additive, convex in outcome mixtures, concave in probabilistic mixtures, and weakly lower semicontinuous, so these properties need not be imposed as axioms. Concavity in probabilistic mixtures captures the aversion to random assignments described above. When two agents are randomly assigned two risks, the sum of their risk measures is at least the corresponding sum when each agent holds one of the risks with certainty. Fair-Coin Compensation then guarantees that the extra risk generated by a fair random assignment can be made arbitrarily small, provided the two institutions can freely trade on the coin. We therefore interpret it as an appealing axiom for regulatory risk measures. Value-at-Risk fails Fair-Coin Compensation (Proposition~\ref{prop:VaR-FC}), whereas Expected Shortfall satisfies it. This distinction provides another reason to prefer Expected Shortfall to Value-at-Risk in financial risk management \citep{embrechts2014}.

The class of OCEs contains many risk measures used in practice. In Section~\ref{sec:selecting}, we characterize several specific members by adding axioms.
The risk-neutral OCE is expectation and satisfies all the axioms considered in this paper. 
We introduce the Equal-Mean Randomization axiom, which characterizes the Mean--Variance criterion (Theorem~\ref{thm:mv}). Equal-Mean Randomization states that selecting one of two losses with the same expected value by a fair coin does not produce a risk measure above the average of their risk measures. The Monotonicity axiom characterizes the standard OCEs (Theorem~\ref{thm:oce}); apart from risk neutrality, it is mutually exclusive with Equal-Mean Randomization.

\begin{figure}[!ht]
    \centering
\begin{tikzpicture}[
  font=\small,
  box/.style={draw=black!65, rounded corners=2pt, fill=black!3,
    line width=.55pt, align=center, inner xsep=7pt, inner ysep=5pt},
  branch/.style={-{Stealth[length=2mm]}, line width=.6pt},
  axiom/.style={font=\footnotesize, inner sep=2pt}
]

\node[box] (oce) at (0,0.4) {\textbf{OCEs}\\(Strong Risk Aversion) + (Normalization) + (Fair-Coin Compensation)\\[1pt]{\color{black!65}Theorem~\ref{thm:main}}};

\node[box] (std) at (-3.0,-1.8) {\textbf{Standard OCEs}\\[1pt]{\color{black!65}Theorem~\ref{thm:oce}}};
\node[box] (mv)  at ( 3.4,-1.8) {\textbf{Mean--Variance}\\[1pt]{\color{black!65}Theorem~\ref{thm:mv}}};

\node[box] (ent) at (-5.2,-3.6) {\textbf{Entropic}\\[1pt]{\color{black!65}Proposition~\ref{prop:entropic}}};
\node[box] (mix) at (0.5,-3.6) {\textbf{Mixtures of $\boldsymbol{\mathbb{E}}$ and Expected Shortfall}\\[1pt]{\color{black!65}Proposition~\ref{prop:homogeneous}}};

\node[box] (es)  at (0.5,-5.4) {\textbf{Expected Shortfall}\\[1pt]{\color{black!65}Corollary~\ref{cor:ES}}};

\draw[branch] (oce) -- node[axiom, left=9pt]  {(Monotonicity)}  (std);
\draw[branch] (oce) -- node[axiom, right=9pt] {(Equal-Mean Randomization)} (mv);
\draw[branch] (std) -- node[axiom, left=9pt]  {(Fair-Coin Independence)} (ent);
\draw[branch] (std) -- node[axiom, right=9pt] {(Positive Homogeneity)} (mix);
\draw[branch] (mix) -- node[axiom, right]       {(Prudence)}  (es);

\end{tikzpicture}
    \caption{Characterizations of OCEs and their subclasses. Each arrow adds the indicated axiom to those already imposed. Within OCEs that are not risk neutral, Monotonicity and Equal-Mean Randomization are mutually exclusive. Within standard OCEs that are not risk neutral, Fair-Coin Independence and Positive Homogeneity are mutually exclusive.}
    \label{fig:placeholder}
\end{figure}
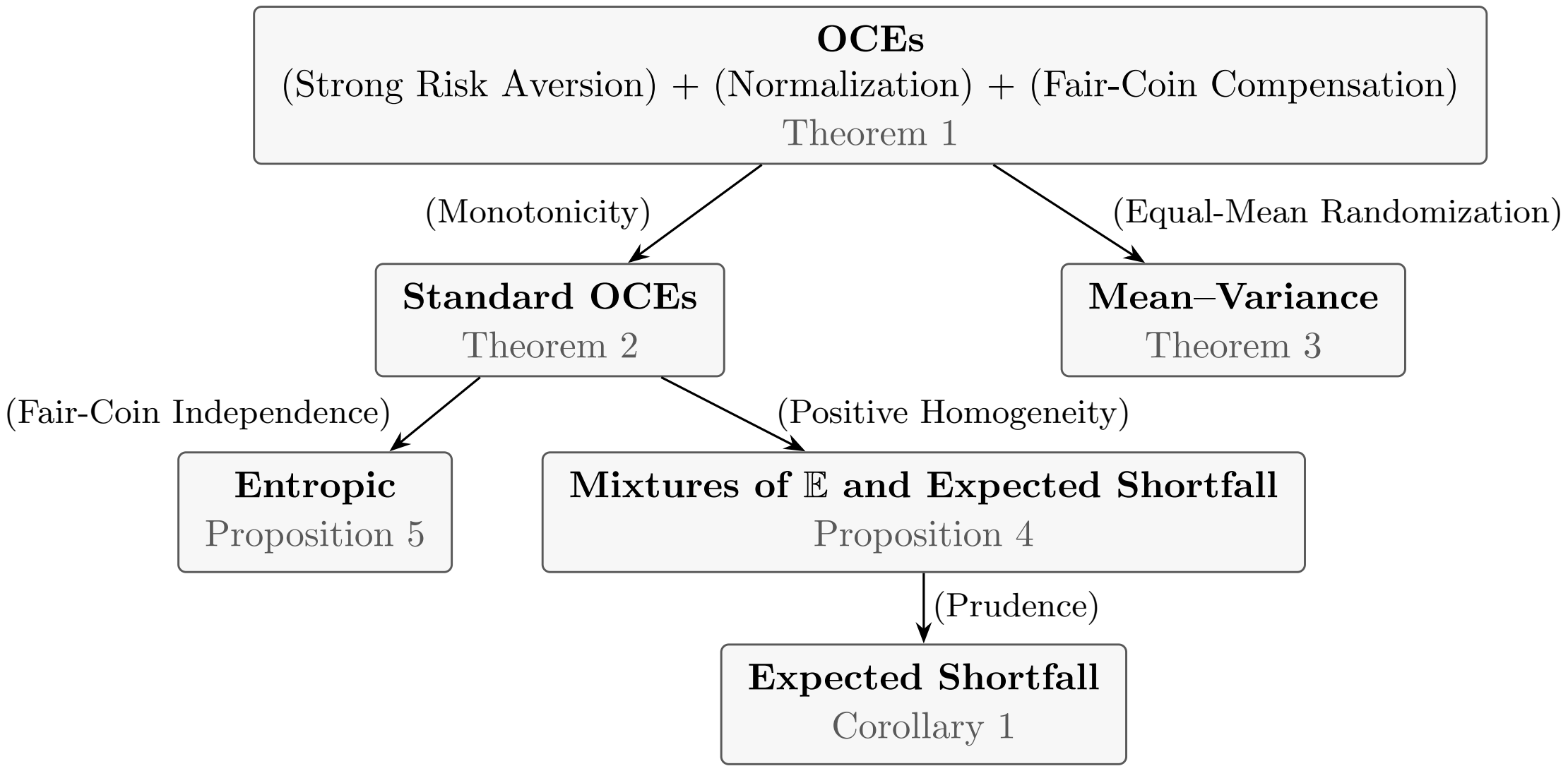

\citet[Proposition 3]{Embrechts2021} showed that positively homogeneous standard OCEs are mixtures of expectation and Expected Shortfall. Together with our main result, this yields an axiomatic characterization of these mixtures (Proposition~\ref{prop:homogeneous}). Adding Prudence \citep{Wang2021, AmaranteLiebrich2024} then yields a novel axiomatic characterization of Expected Shortfall (Corollary~\ref{cor:ES}). Finally, the Fair-Coin Independence axiom characterizes the Entropic Risk Measure (Proposition~\ref{prop:entropic}).  Fair-Coin Independence requires that mixing two losses with a common third loss via a fair coin does not change their ranking; it is also known as the Independent Sure Thing Principle in decision theory. Fair-Coin Independence is mutually exclusive with Positive Homogeneity for risk measures that are not risk neutral (that is, not the expectation). We summarize all these relationships in Figure~\ref{fig:placeholder}.

\subsection{Related literature}

Axiomatic approaches to risk measures have a long tradition, starting with the coherent risk measures of \citet{Artzner1999} and the convex risk measures of \citet{Follmer2002} and \citet{Frittelli2002}, and with \citet{Kusuoka2001} representing law-invariant coherent risk measures through the Expected Shortfall. 

Some axiomatic characterizations are closely related to our results. \citet{Wang2021} characterize the Expected Shortfall through four axioms: Monotonicity, Law Invariance, Prudence, and No Reward for Concentration. Their axioms do not include subadditivity, comonotonic additivity, or the Lebesgue property, and thus, these properties are implied by Prudence and No Reward for Concentration. Similarly, our characterization of the Expected Shortfall does not assume any of these properties but derives them. \cite{AmaranteLiebrich2024} provides an alternative analysis that clarifies the role of the Prudence axiom.

Our characterization of Mean--Expected Shortfall builds on \citet{Embrechts2021}, who characterized positively homogeneous standard OCEs. Their characterization, however, assumes an OCE representation rather than deriving it from axioms on the risk measure. \citet{Embrechts2021} also characterize Expected Shortfall as the only coherent Bayes risk measure and Entropic Risk Measures as the only risk measures that are both elicitable and Bayes. \citet{Han2024} further analyzes the Mean--Expected Shortfall criterion. 

\citet{Pomatto2020} characterizes the Mean--Variance class as the unique class of finite-valued functionals satisfying Normalization, Strong Risk Aversion, and additivity for independent random variables on a space admitting unbounded random variables. However, contrary to our axiomatization, their axioms do not pin down the Mean--Variance criterion on $L^\infty$ because the Entropic Risk Measure also satisfies them on this space. We therefore provide the first characterization of Mean--Variance, to our knowledge, on $L^\infty$. 

The standard OCE was introduced by \citet{BenTal1986} and connected with $\varphi$-divergence penalties by \citet{BenTal1987}. \citet{BenTal2007} studied its properties as a convex risk measure, derived its dual representation, and showed that, among standard OCEs with finite loss functions satisfying $\ell(x)>x$ for every $x\ne0$, the coherent ones are generated by piecewise linear loss functions. The $\varphi$-divergences were introduced by \citet{Csiszar1967}, and \citet{Liese1987} study their properties. These papers start from the representation \eqref{eq:oce} or from its dual form, whereas we derive this representation from axioms on the risk measure. 

In decision theory, the dual form of standard OCEs corresponds to the divergence preferences that \citet{Maccheroni2006} obtain as a special case of variational preferences. Their axioms apply to preferences over acts in the Anscombe--Aumann framework and characterize the whole class of variational preferences, and divergence preferences enter as a special case without axioms of their own. Our axioms apply to law-invariant risk measures with known distributions, and Fair-Coin Compensation identifies the OCEs among them. 
\cite{WuMaoHu2024} generalize optimized certainty equivalents to incorporate ambiguity through variational preferences. \citet{Sarver2018} represents mixture-averse preferences as a maximum over a family of expected utilities, and standard OCEs correspond to the case where this family consists of translations of a single utility function.
\citet{Strzalecki2011} characterize the multiplier preferences of \citet{Hansen2001}, which penalize alternative models by relative entropy, by imposing the sure-thing principle of \citet{Savage1954} to variational preferences. 

Several strong consistency requirements characterize the Entropic Risk Measure within the class of standard OCEs. Examples are recursive time consistency under law invariance \citep{Kupper2009} and the combination of concavity in probabilistic mixtures with convex level sets \citep{Embrechts2021}. Since the Entropic Risk Measure is one specific OCE, characterizing the whole class must rely on weaker axioms than those of the Entropic Risk Measure.

OCEs are concave in probabilistic mixtures. \citet{Acciaio2013} analyze when law-invariant convex risk measures are concave in probabilistic mixtures, and \citet{Delage2019} examine how this property relates to when randomized decisions can reduce risk. In our setting, the axioms imply concavity in probabilistic mixtures and underlie our interpretation of Fair-Coin Compensation.

The remainder of this paper is structured as follows. We define OCEs and give their main examples in Section~\ref{sec:oce}. In Section~\ref{sec:axioms}, we introduce the axioms, derive their consequences, interpret Fair-Coin Compensation, and state the characterization. Section~\ref{sec:selecting} characterizes standard OCEs, Mean--Variance risk measures, Expected Shortfall, and Entropic Risk Measures by additional properties. We conclude in Section~\ref{sec:conclusion}. We defer the proofs to the appendices.

\section{Optimized certainty equivalents}\label{sec:oce}

We first fix the notation, then define OCEs via convex loss functions, and present the examples we use throughout the paper. We define all random variables in this paper on a probability space $(\Omega,\cF,\P)$.

\begin{assumption}\label{ass:atomless}
The probability space $(\Omega,\cF,\P)$ is atomless.
\end{assumption}

Under Assumption~\ref{ass:atomless}, we can construct a random vector on $\Omega$ with any prescribed distribution on $\R^n$ \citep[Proposition~A.31]{Follmer2016}, a fact we use often in the proofs.

The space $L^\infty$ is the set of essentially bounded random variables on $(\Omega,\cF,\P)$, where we identify random variables that are equal almost surely. A random variable $X\in L^\infty$ represents a loss at a fixed horizon, so that positive values are losses and negative values are gains. Equalities and inequalities between random variables are understood in the $\P$-almost-sure sense. We write $X\eqd Y$ if $X$ and $Y$ have the same distribution under $\P$ and $\|X\|_\infty$ for the sup-norm on $L^\infty$. We denote by $\mathcal D^\infty$ the set of distributions on $\mathbb{R}$ with bounded support. Given a random variable $X$, we write $\mu_X$ for the distribution of $X$, that is, $\mu_X(A)=\P(X\in A)$ for every Borel set $A\subseteq\R$. We write $X\lecx Y$, and say that $X$ is smaller than $Y$ in the convex order, if $\E[\phi(X)]\le\E[\phi(Y)]$ for every convex function $\phi:\R\to\R$.
We denote by $\sigma(L^\infty,L^1)$ the coarsest topology on $L^\infty$ that makes the map $X\mapsto\E[QX]$ continuous for every $Q\in L^1$, where $L^1$ is the space of integrable random variables on $(\Omega,\cF,\P)$.
For $x\in\R$, we write $x_+=\max\{x,0\}$.

A \emph{coin with probability $\lambda\in(0,1)$} is a random variable $B$ with $\P(B=1)=\lambda$ and $\P(B=0)=1-\lambda$, and a fair coin is a coin with probability $1/2$. If $B$ is a coin with probability $\lambda$ that is independent of $(X,Y)$, then
$BX+(1-B)Y$ has distribution $\lambda\mu_X+(1-\lambda)\mu_Y$.
This \emph{probabilistic mixture} is the distribution defined by
$$
\bigl(\lambda\mu_X+(1-\lambda)\mu_Y\bigr)(A)
=\lambda\mu_X(A)+(1-\lambda)\mu_Y(A)
$$
for every Borel set $A\subseteq\R$.

A \emph{risk measure} is a mapping $\rho: L^\infty \to \R$, and $\rho(X)$ is interpreted as the capital requirement of the loss $X$. We use the following properties of a risk measure $\rho$.
\begin{itemize}
\item[] Monotonicity: $\rho(X)\le\rho(Y)$ whenever $X\le Y$.
\item[] Law invariance: $\rho(X)=\rho(Y)$ whenever $X\eqd Y$.
\item[] Cash additivity: $\rho(X+c)=\rho(X)+c$ for all $X\in L^\infty$ and $c\in\R$.
\item[] Convexity in outcome mixtures:  $\rho(\lambda X+(1-\lambda)Y)\le\lambda\rho(X)+(1-\lambda)\rho(Y)$ for all $X,Y\in L^\infty$ and $\lambda\in[0,1]$.
\item[] Weak lower semicontinuity: $\{X\in L^\infty:\rho(X)\le c\}$ is closed in $\sigma(L^\infty,L^1)$ for every $c\in\R$.
\end{itemize}
We say that a risk measure $\rho$ is a \emph{law-invariant convex risk measure} if it is monotone, law invariant, cash additive, convex in outcome mixtures, and satisfies $\rho(0)=0$. For a law-invariant $\rho$, we write $\widehat\rho:\mathcal D^\infty\to\R$ for the functional with $\widehat\rho(\mu_X)=\rho(X)$. We say that a law-invariant risk measure $\rho$ is \emph{concave in probabilistic mixtures} if $$\widehat\rho(t\mu+(1-t)\nu)\ge t\widehat\rho(\mu)+(1-t)\widehat\rho(\nu)$$ for all $\mu,\nu\in\mathcal D^\infty$ and $t\in[0,1]$.

The two mixture properties correspond to two different ways of mixing two losses. The outcome mixture $\lambda X+(1-\lambda)Y$ combines the two losses in every state, so a risk measure convex in outcome mixtures promotes diversification. The probabilistic mixture $BX+(1-B)Y$ selects one of the two losses at random, so a law-invariant risk measure concave in probabilistic mixtures does not reward random assignment.

\begin{definition}\label{def:loss}
A \emph{loss function} is a convex, lower semicontinuous function $\ell:\R\to\R\cup\{+\infty\}$ with an unbounded domain such that $\ell(0)=0$ and $\ell(x)\ge x$ for every $x\in\R$. We denote by $\cL$ the set of loss functions and by $\cL_+$ the set of nondecreasing loss functions. Two loss functions $\ell$ and $\ell'$ are \emph{equivalent} if $\ell'(x)=\ell(x+c)-c$ for some $c\in\R$ and every $x\in\R$.
\end{definition}

The condition $\ell(x)\ge x$ means that a residual loss is penalized at least at its face value, and the other conditions ensure that the risk measure below is finite.

\begin{definition}\label{def:oce}
For $\ell\in\cL$, the \emph{OCE} $\rho_\ell$ is
$$
\rho_\ell(X)=\inf_{m\in\R}\left\{m+\E[\ell(X-m)]\right\},~~X\in L^\infty.
$$
If $\ell\in\cL_+$, we call $\rho_\ell$ a \emph{standard OCE}, as in \citet{BenTal1986,BenTal2007}.
\end{definition}

The assumptions on $\ell$ ensure that $\rho_\ell$ is real valued on
$L^\infty$. Indeed, $\ell(x)\ge x$ bounds the objective below by
$\E[X]$, while the boundedness of $X$ and the unbounded interval on
which $\ell$ is finite allow a cash amount $m$ for which
$\ell(X-m)$ is bounded. \citet{BenTal1986, BenTal2007} interpret a standard OCE as the present cost of a loss to a decision maker who can pay part of it in advance: paying $m$ now leaves the residual loss $X-m$, whose disutility is $\E[\ell(X-m)]$, and $\rho_\ell(X)$ is the smallest total cost that the decision maker can achieve by choosing $m$.

The following examples are common classes of OCEs found in the literature.

\begin{example}\label{ex:MV}
For $\kappa\ge0$, the loss function $\ell(x)=x+\kappa x^2$ gives
$$
\rho_\ell(X)
=\E[X]+\kappa\inf_{m\in\R}\E[(X-m)^2]
=\E[X]+\kappa\Var(X),~~X\in L^\infty.
$$
This recovers the \emph{Mean--Variance} criterion used in portfolio selection
\citep{Markowitz1952}. Since $\ell$ is quadratic when $\kappa>0$, $\ell\notin \mathcal{L}_+$ when $\kappa>0$. Therefore, Mean--Variance is an OCE, but not a standard OCE, when $\kappa>0$.
\end{example}

\begin{example}\label{ex:ES}
For $p\in(0,1)$, the nondecreasing loss function
$\ell(x)=x_+/(1-p)$ gives the \emph{Expected Shortfall} at level $p$.
By the minimization formula of
\citet{Rockafellar2000,Rockafellar2002},
$$
\rho_\ell(X)
=\inf_{m\in\R}\left\{m+\frac{1}{1-p}\E[(X-m)_+]\right\}
=\ES_p(X),~~X\in L^\infty,
$$
where
$$
\ES_p(X)=\frac{1}{1-p}\int_p^1\VaR_s(X)\,\mathrm ds,~~
\VaR_s(X)=\inf\{x\in\R:\P(X\le x)\ge s\},~~X\in L^\infty,
$$
and $\VaR_s$ is the \emph{VaR at level $s$}.
Since $\ell\in\cL_+$, Expected Shortfall is a standard OCE. We also write $\ES_1=\esssup$. This is the standard OCE generated
by the loss function that equals $0$ on $(-\infty,0]$ and
$+\infty$ on $(0,\infty)$.
\end{example}

\begin{example}\label{ex:entrop}
For $\gamma>0$, the nondecreasing loss function
$\ell(x)=(e^{\gamma x}-1)/\gamma$ gives the \emph{Entropic Risk Measure}
\begin{equation}\label{eq:entropic}
\mathrm{ER}_\gamma(X)
=\rho_\ell(X)
=\frac{1}{\gamma}\log\E[e^{\gamma X}].
\end{equation}
Thus the Entropic Risk Measure is a standard OCE.
\end{example}

Every OCE is law invariant, cash additive, convex in outcome
mixtures, concave in probabilistic mixtures, and weakly lower
semicontinuous.
We verify these properties
in Lemma~\ref{lem:oce-penalty-working} and in the proofs of
Theorems~\ref{thm:main} and~\ref{thm:oce} in Appendix~\ref{app:A}.

\section{Axioms and the characterization}\label{sec:axioms}

This section presents our axiomatic characterization of optimized certainty equivalents. We first introduce the main axioms, derive their consequences, and then interpret the new Fair-Coin Compensation axiom. Theorem~\ref{thm:main} shows that the three axioms characterize OCEs, and Theorem~\ref{thm:oce} shows that adding Monotonicity characterizes standard OCEs. We conclude by showing that the axioms are independent.

\subsection{The axioms}\label{sec:standard}

The first two axioms are standard.

\begin{axiom}[(SRA) (Strong Risk Aversion)]
A loss that is smaller in the convex order does not have a larger risk evaluation. That is, $\rho(X)\le\rho(Y)$ whenever $X\lecx Y$.
\end{axiom}

Axiom~(SRA) compares a loss with a more variable loss. If $X\eqd Y$, then $X\lecx Y$ and $Y\lecx X$, so law invariance is a necessary condition for Axiom~(SRA). On an atomless probability space, every law-invariant convex risk measure satisfies Axiom~(SRA); see \citep[Theorem~30]{Delbaen2020} and \citep[Theorem~3.6]{Bellini2021}.

\begin{axiom}[(N) (Normalization)]
The risk evaluation of a sure loss is its amount. That is, $\rho(c)=c$ for every $c\in\R$.
\end{axiom}

Axiom~(N) makes $\rho(X)$ the certainty equivalent of $X$: the
institution is indifferent between facing the uncertain loss $X$
and paying the sure amount $\rho(X)$, since
$\rho(\rho(X))=\rho(X)$.

The third axiom involves a random assignment of one of two losses. Let $B$ be a fair coin that is independent of $(X,Y)$. For $a\in\R$, the loss $B(X+a)+(1-B)(Y-a)$ adds $a$ to the loss when $X$ is selected and subtracts $a$ when $Y$ is selected, with the loss selected based on the coin's outcome. Since this adjustment depends only on the coin's outcome, we call it \emph{a transfer on the coin}.

\begin{axiom}[(FC) (Fair-Coin Compensation)]
For two losses $X$ and $Y$ and a fair coin that is independent of $(X, Y)$, the smallest risk measure that transfers on the coin can achieve is the average of the risk measures of $X$ and $Y$. That is,
$$
\inf_{a\in\R}\rho\left(B(X+a)+(1-B)(Y-a)\right)=\frac{\rho(X)+\rho(Y)}{2}
$$
for all $X,Y\in L^\infty$ and every fair coin $B$ that is independent of $(X,Y)$.
\end{axiom}

We interpret this axiom in Section~\ref{sec:FC}. The next result collects the properties implied by the three axioms.

\begin{proposition}\label{prop:properties}
Let $\rho: L^\infty\to\R$ satisfy Axioms~(SRA), (FC), and (N). Then $\rho$ is law invariant, cash additive, convex in outcome mixtures, concave in probabilistic mixtures, and weakly lower semicontinuous.
\end{proposition}

Proposition~\ref{prop:properties} does not include monotonicity, which we state as a separate axiom.

\begin{axiom}[(M) (Monotonicity)]
A surely larger loss does not have a smaller risk measure. That is, $\rho(X)\le\rho(Y)$ whenever $X\le Y$.
\end{axiom}

Axiom~(M) is standard in risk measure theory, and every standard OCE satisfies it, as verified in Theorem~\ref{thm:oce}. We do not impose Axiom~(M) in the main characterization of Section~\ref{sec:main}, because the Mean--Variance risk measures presented in Example~\ref{ex:MV} are not standard OCEs.

\subsection{Fair assignment and transfers on the coin}\label{sec:FC}

Axioms~(SRA) and (N) are familiar in the theory of risk measures, whereas Axiom~(FC) is new. Axiom~(FC) has a natural economic interpretation in terms of the fair assignment of losses, a well-accepted mechanism in economics.

Let $\rho$ be a risk measure that satisfies Axioms~(N) and~(SRA), is concave in probabilistic mixtures, and is weakly lower semicontinuous. Many familiar risk measures satisfy these properties. Besides OCEs, examples include certainty equivalents of expected loss with convex loss, see \cite{Savage1954}; distortion risk measures with concave distortion functions, see \citet{Acerbi2002}, which include Expected Shortfall; the worst-case expectations over divergence-based balls, see \cite{BreuerCsiszar2016}, which include the Entropic Value-at-Risk of \citet{AhmadiJavid2012}.

Consider two financial institutions that evaluate risk with the risk measure $\rho$, for example because a regulator prescribes it. Additionally, assume the same financial regulator must assign losses $X$ and $Y$ to these two participants, where $\rho(X)<\rho(Y)$. Assume the losses cannot be split ex ante, and each participant is assigned one loss. Therefore, assigning $Y$ to one participant deterministically favours the other and leads to an inequitable division. 

Instead, the two losses can be assigned by a fair coin $B$ that is independent of $(X, Y)$, so that the losses of the two participants are
$$W_1(0)=BX+(1-B)Y~~\text{and}~~ W_2(0)=BY+(1-B)X.$$
Both participants have the same chance of receiving either loss, and they face the same loss distribution before the coin is tossed. \citet{Fudenberg2012} discuss this notion of fairness, in which a random allocation treats individuals equally before the outcome is realized. Such random allocations are used in practice \citep{Budish2013}, for example, to break ties between the highest bids in auctions \cite{myerson1981}, and their effect on participants' welfare depends on how they evaluate probabilistic mixtures \citep{Dillenberger2021, Dillenberger2025}. 

Since $\rho$ is concave in probabilistic mixtures,
$$\rho\left(W_1(0)\right)+\rho\left(W_2(0)\right)\ge\rho(X)+\rho(Y).$$
We call the difference between the two sides the \emph{fairness premium} of the fair assignment. It is the additional aggregate risk measure, that is, the additional sum of the two participants' risk measures, that results from assigning the losses at random. The fairness premium is nonnegative and can be zero. The total loss $W_1(0)+W_2(0)=X+Y$ is the same in every state, so fairness means equal treatment before the coin is tossed.

The random assignment exposes each participant to the risk of receiving the less desirable $Y$. To reduce this risk, before the coin is tossed, the two participants can agree on a transfer on the coin. Their losses are then
$$W_1(a)=B(X+a)+(1-B)(Y-a)~~\text{and}~~W_2(a)=(1-B)(X+a)+B(Y-a).$$
We have $W_1(a)+W_2(a)=X+Y$, so the transfer does not change the aggregate loss.

Since the coin is fair and independent of $(X, Y)$, the two
participants face the same loss distribution for every transfer
$a$. By law invariance, their aggregate risk is therefore
$$R(a)=\rho(W_1(a))+\rho(W_2(a))=2\rho(W_1(a)).$$
We define the \emph{fairness premium} associated with the
transfer $a$ by
$$\pi_\rho(a)=R(a)-\rho(X)-\rho(Y).$$
Axiom~(FC) requires that
$\inf_{a\in\R}R(a)=\rho(X)+\rho(Y),$
so that the fairness premium can be made arbitrarily close
to zero. The following lemma ensures that the infimum is
attained under our standing assumptions.

\begin{lemma}\label{lem:ref}
Let $\rho$ satisfy Axiom~(SRA) and be weakly lower semicontinuous. Then, for every $X,Y\in L^\infty$, there exists
$a^*\in\R$ such that
$
R(a^*)=\inf_{a\in\R}R(a).
$
\end{lemma}

Consequently, under Axiom~(FC), the optimal transfer $a^*$
achieves $\pi_\rho(a^*)=0$: the participants can agree on a
fair assignment without increasing aggregate risk.
Moreover, both participants weakly prefer this arrangement
to the uncompensated fair assignment, since
$$
\rho(W_i(a^*))=\frac12 R(a^*)
\le \frac12 R(0)=\rho(W_i(0)),
\qquad i=1,2.
$$
The improvement is strict for both whenever $\pi_\rho(0)>0$.
Thus, Axiom~(FC) ensures that any additional aggregate risk
arising from the fair assignment can be eliminated through
an appropriate transfer.

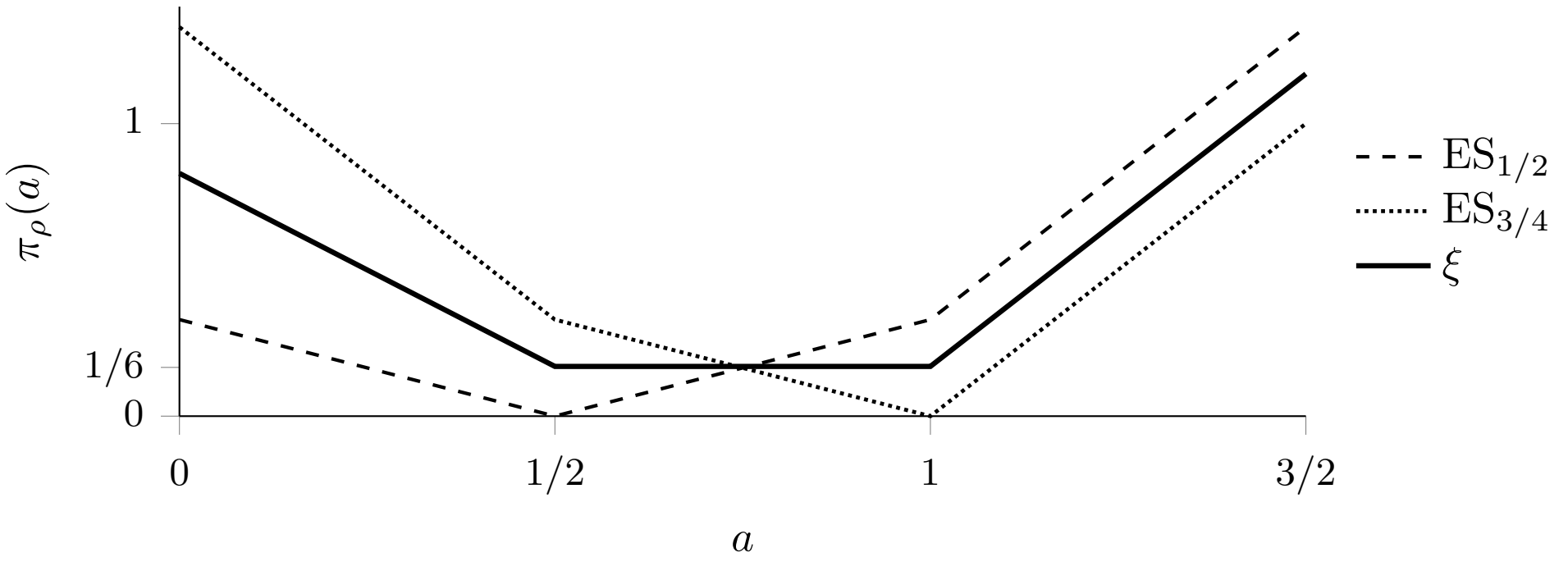
\begin{figure}[!ht]
\centering
\begin{tikzpicture}
\begin{axis}[paperplot,width=0.55\textwidth,height=0.2\textwidth,
  xmin=0,xmax=1.5,ymin=0,ymax=1.4,
  xtick={0,0.5,1,1.5},xticklabels={$0$,$1/2$,$1$,$3/2$},
  ytick={0,0.166667,1},yticklabels={$0$,$1/6$,$1$},
  xlabel={$a$},ylabel={$\pi_\rho(a)$},
  legend style={at={(1.03,0.5)},anchor=west,draw=none,font=\small},legend cell align=left]
\addplot[thick,dashed] coordinates {(0,0.33) (0.5,0) (1,0.33) (1.5,1.33)};
\addlegendentry{$\ES_{1/2}$}
\addplot[line width=0.9pt,densely dotted] coordinates {(0,1.33) (0.5,0.33) (1,0) (1.5,1)};
\addlegendentry{$\ES_{3/4}$}
\addplot[very thick] coordinates {(0,0.83) (0.5,0.17) (1,0.17) (1.5,1.17)};
\addlegendentry{$\xi$}
\end{axis}
\end{tikzpicture}
\caption{Fairness premium as a function of the transfer.}\label{fig:premium}
\end{figure}

\begin{example}\label{ex:premium}
Let $X=0$, and let $Y$ be uniformly distributed on $\{0,1,2\}$. Figure~\ref{fig:premium} plots the fairness premium $\pi_\rho(a)$ against the transfer $a$ for $\ES_{1/2}$, $\ES_{3/4}$, and their average $\xi=\frac12\ES_{1/2}+\frac12\ES_{3/4}$. The premium of $\ES_{1/2}$ reaches zero at transfer $1/2$, and the premium of $\ES_{3/4}$ reaches zero at transfer $1$.

Since $\xi$ is the average of the two ES, the premium of $\xi$ at every transfer is therefore the average of the premiums of the two ES. The two ES have zero premiums at different transfers, so $\xi$ has a positive premium at every transfer. Its smallest value is $1/6$, attained at every transfer between $1/2$ and $1$, so $\xi$ violates Axiom~(FC), although it is convex in outcome mixtures, concave in probabilistic mixtures, and weakly lower semicontinuous.
\end{example}

Axiom~(FC) applies to any risk measure, including VaR, which is not an OCE. Given VaR's role in financial regulation, it is worth asking whether it satisfies Axiom~(FC).
\begin{proposition}\label{prop:VaR-FC}
For every $p\in(0,1)$, $\VaR_p$ does not satisfy Axiom~(FC).
\end{proposition}
If one believes Axiom~(FC) is desirable for regulatory risk management, then Proposition \ref{prop:VaR-FC} provides another rationale for preferring ES over VaR.

\subsection{Characterization of OCEs}\label{sec:main}

We now state the main result of this paper, which characterizes OCEs by the three axioms.

\begin{theorem}\label{thm:main}
A risk measure $\rho: L^\infty \to \R$ satisfies Axioms~(SRA), (FC), and (N) if and only if $\rho$ is an OCE, that is, $\rho = \rho_ \ell$ for some $\ell \in \cL$. Moreover, if $\rho \ne \E$, then for every loss function $\ell' \in \cL$, we have $\rho_{\ell'}=\rho$ if and only if $\ell'$ is equivalent to $\ell$.
\end{theorem}

The class in Theorem~\ref{thm:main} consists of the OCEs generated by all loss functions in $\cL$, which include standard OCEs and the Mean--Variance criterion presented in Example~\ref{ex:MV}. Moreover, Theorem~\ref{thm:main} extends a similar uniqueness result from \citet{BenTal2007} to the larger class $\cL$. 

By Theorem~\ref{thm:main}, if $\ell(x)>x$ for every $x\ne0$, then $\ell$ is the only loss function that generates $\rho_\ell$. 

We now axiomatize the class of standard OCEs, as \citet{BenTal1986} originally formulated. The next result shows that the standard OCEs are the risk measures that satisfy the three axioms of Theorem~\ref{thm:main} and Axiom~(M).

\begin{theorem}\label{thm:oce}
A risk measure $\rho: L^\infty \to \R$ satisfies Axioms~(SRA), (FC), (N), and (M) if and only if $\rho$ is a standard OCE, that is, $\rho = \rho_ \ell$ for some $\ell \in \cL_+$.
\end{theorem}

The next proposition shows that none of the four axioms of Theorem~\ref{thm:oce} can be dropped.

\begin{proposition}\label{prop:indep}
Axioms~(SRA), (FC), (N), and (M) are independent.
\end{proposition}

Here independence means that each axiom can fail while the other three hold. 

\begin{figure}[!ht]
\centering
\begin{tikzpicture}[scale=1.15, every node/.style={font=\small}]
\def\sraPath{[rotate around={50:(-.3,.2)}](-.3,.2) ellipse (2.5 and 1.6)}
\def\fcPath{[rotate around={-50:(.3,.2)}](.3,.2) ellipse (2.5 and 1.6)}
\def\nPath{[rotate around={-36:(-.6,-.5)}](-.6,-.5) ellipse (2.5 and 1.6)}
\def\mPath{[rotate around={36:(.6,-.5)}](.6,-.5) ellipse (2.5 and 1.6)}
\begin{scope}
\clip \sraPath; \clip \fcPath; \clip \nPath;
\fill[blue!13] (-4,-4) rectangle (4,4);
\clip \mPath;
\fill[blue!32] (-4,-4) rectangle (4,4);
\end{scope}
\draw[blue!65!black, line width=.9pt] \sraPath;
\draw[teal!70!black, line width=.9pt] \fcPath;
\draw[orange!80!black, line width=.9pt] \nPath;
\draw[violet!75!black, line width=.9pt] \mPath;
\node[blue!65!black, anchor=south west] at (1.31,2.12) {(SRA)};
\node[teal!70!black, anchor=south east] at (-1.31,2.12) {(FC)};
\node[orange!80!black, anchor=east] at (-2.68,0.97) {(N)};
\node[violet!75!black, anchor=west] at (2.68,0.97) {(M)};
\node[align=center] at (0,-.28) {Standard\\OCEs};
\fill (-1.12,.86) circle (1.4pt) node[above=1pt] {$\rho_{\mathrm{MV}}$};
\fill (1.12,.86) circle (1.4pt) node[above=1pt] {$2\mathbb E$};
\fill (-1.08,-1.44) circle (1.4pt) node[above=1pt] {$\xi$};
\fill (1.08,-1.44) circle (1.4pt) node[above=1pt] {$\chi$};
\begin{scope}[shift={(4.1,-.1)}]
\fill[blue!13] (0,.17) rectangle (.25,.33);
\draw[line width=.3pt] (0,.17) rectangle (.25,.33);
\node[anchor=west] at (.35,.25) {OCEs};
\fill[blue!32] (0,-.33) rectangle (.25,-.17);
\draw[line width=.3pt] (0,-.33) rectangle (.25,-.17);
\node[anchor=west] at (.35,-.25) {Standard OCEs};
\end{scope}
\end{tikzpicture}
\caption{Independence of Axioms~(SRA), (FC), (N), and (M). Each marked risk measure satisfies the three axioms whose sets contain it and violates the remaining axiom. The light- and dark-shaded regions form the OCEs; the dark region forms the standard OCEs.}\label{fig:venn}
\end{figure}
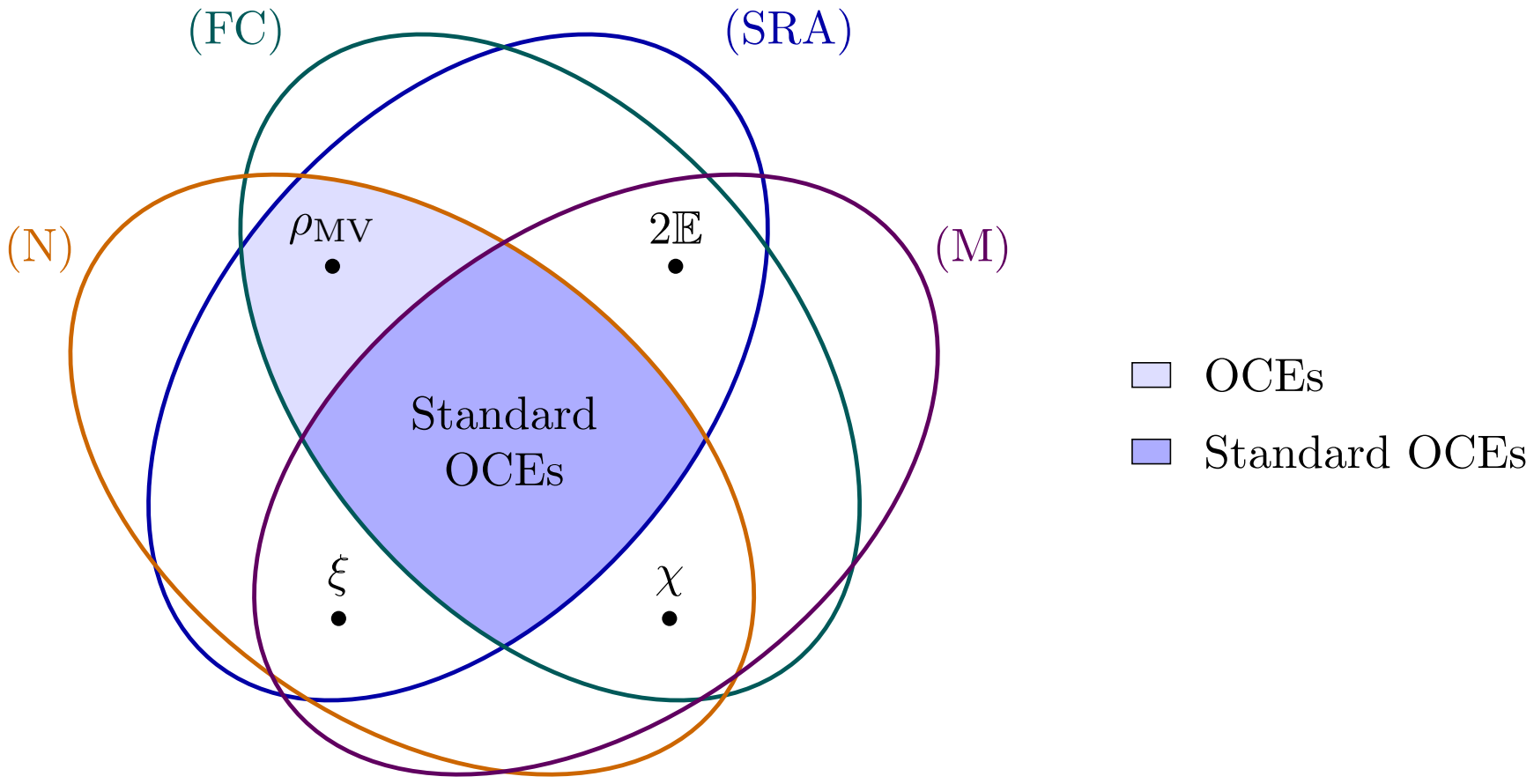

We verify Proposition~\ref{prop:indep} in Appendix~\ref{app:A} using four risk measures, each satisfying three axioms and violating the fourth, as illustrated in Figure~\ref{fig:venn}. The first is the average $\xi=\frac12\ES_{1/2}+\frac12\ES_{3/4}$ of Example~\ref{ex:premium}, which satisfies Axioms~(SRA), (N), and (M) but violates Axiom~(FC). 
 
The second is the risk measure
$$\chi(X)=\inf_{m\in\R}\left\{m+\E[(X-m)_+]+\P(X>m)\right\},$$
which has the form \eqref{eq:oce} with the nondecreasing but nonconvex loss function $x_++\mathbbm{1}_{\{x>0\}}$. The risk measure $\chi$ satisfies Axioms~(FC), (N), and (M) but violates Axiom~(SRA). 

The third is the risk measure $2\E$, which satisfies Axioms~(SRA), (FC), and (M) but fails to satisfy Axiom~(N).
The fourth is the Mean--Variance criterion in Example~\ref{ex:MV}, an OCE that satisfies Axioms~(SRA), (FC), and (N) but violates Axiom~(M). In Figure~\ref{fig:venn}, the OCEs form the region where the sets of Axioms~(SRA), (FC), and (N) overlap, and the standard OCEs form the part of this region that also satisfies Axiom~(M).

\section{Characterizations of specific OCEs}\label{sec:selecting}

Theorem~\ref{thm:main} characterizes a family of risk measures, and the choice of loss function determines the risk measure within the family. We first characterize the Mean--Variance criterion through Equal-Mean Randomization. Sections~\ref{sec:homogeneity} and~\ref{sec:entropic} then combine our representation results with established classifications to study Positive Homogeneity, Prudence, and Fair-Coin Independence.

\subsection{The Mean--Variance criterion}\label{sec:ER}

We characterize the Mean--Variance criterion through the following property.

\begin{property}[(ER) (Equal-Mean Randomization)]
Selecting one of two losses with the same expected value by a fair coin does not produce a risk measure above the average of their risk measures. That is,
$$\rho\left(BX+(1-B)Y\right)\le\frac{\rho(X)+\rho(Y)}{2}$$
for all $X,Y\in L^\infty$ with $\E[X]=\E[Y]$ and every fair coin $B$ that is independent of $(X,Y)$.
\end{property}

If $\rho$ satisfies Axioms~(SRA), (N), and (FC), then Theorem~\ref {prop:properties} implies that $\rho$ is concave in probabilistic mixtures. Therefore, Property~(ER), in combination with these axioms, implies that $\rho$ is midpoint affine in mixtures of distributions with a common mean, that is, for all $X,Y\in L^\infty$ with $\E[X]=\E[Y]$ and every fair coin $B$ that is independent of $(X,Y)$, we have
$$
\rho(BX+(1-B)Y)=\frac{\rho(X)+\rho(Y)}{2}.
$$
Equivalently, the zero transfer attains the infimum in Axiom~(FC) whenever $\E[X]=\E[Y]$.

\begin{theorem}\label{thm:mv}
A risk measure $\rho: L^\infty \to \R$ satisfies Axioms~(SRA), (FC), and (N) and Property~(ER) if and only if there exists $\kappa\ge0$ such that
$$\rho(X)=\E[X]+\kappa\operatorname{Var}(X),~~X\in L^\infty.$$
\end{theorem}

The quadratic loss function $x+\kappa x^2$ is nondecreasing only for $\kappa=0$, so by Theorem~\ref{thm:oce}, the expectation is the only standard OCE that satisfies Property~(ER).

\subsection{Expected Shortfall}\label{sec:homogeneity}

Standard OCEs represented by a finite-valued loss function $\ell\in \mathcal{L}_+$ belong to the normalized monetary convex Bayes class of \citet[Remark~3.3]{Embrechts2021}. Therefore, similar to their work, we consider the following property.

\begin{property}[(PH) (Positive Homogeneity)]
Scaling a loss by a positive factor scales its risk measure by the same factor. That is, $\rho(tX)=t\rho(X)$ for all $X\in L^\infty$ and $t>0$.
\end{property}

By \citet[Theorem~3.1]{Embrechts2021}, if a Standard OCE is represented by a finite-valued loss function $\ell\in \mathcal{L}_+$, then the following result holds. However, we provide a proof for the extended-valued case, which supplies the endpoint $\mathrm{ES}_1=\operatorname{ess\,sup}$.

\begin{proposition}\label{prop:homogeneous}
A risk measure $\rho: L^\infty \to \R$ satisfies Axioms~(SRA), (FC), (N), and (M) and Property~(PH) if and only if there exist $\lambda\in[0,1]$ and $p\in(0,1]$ such that
$$\rho(X)=\lambda\E[X]+(1-\lambda)\mathrm{ES}_p(X),~~X\in L^\infty.$$
\end{proposition}

Proposition~\ref{prop:homogeneous} characterizes the mixtures $\lambda\E+(1-\lambda)\ES_p$. To isolate $\ES_p$ within this family, we impose Prudence, the axiom introduced in \cite{Wang2021}.

\begin{property}[(P) (Prudence)]
The risk measure of an almost-sure limit of losses does not exceed the limit inferior of their risk measures. That is, for every sequence $(X_n)_{n\in\mathbb N}\subset L^\infty$ and $X\in L^\infty$ such that $\lim_{n\to\infty}X_n=X$ $\P$-almost surely, we have
$$\liminf_{n\to\infty}\rho(X_n)\ge\rho(X).$$
\end{property}

\begin{corollary}\label{cor:ES}
A risk measure $\rho: L^\infty \to \R$ satisfies Axioms~(SRA), (FC), (N), and (M) and Properties~(PH) and (P) if and only if there exists $p\in(0,1]$ such that
$$\rho(X)=\mathrm{ES}_p(X),~~X\in L^\infty.$$
\end{corollary}

Therefore, Corollary~\ref{cor:ES} provides an alternative characterization of Expected Shortfall.

\subsection{Entropic Risk Measures}\label{sec:entropic}

We now introduce our final property, Fair-Coin Independence, and use it to characterize the Entropic Risk Measures. This property concerns how risk evaluations respond when losses are mixed with a common third loss by an independent fair coin.

\begin{property}[(FI) (Fair-Coin Independence)]
Mixing each of two losses with a common third loss through a fair coin does not change their ranking. That is,
$$
\rho(X)\le\rho(Y)
~~\text{if and only if}~~
\rho\left(BX+(1-B)Z\right)
\le
\rho\left(BY+(1-B)Z\right)
$$
for all $X,Y,Z\in L^\infty$ and every fair coin $B$ that is independent of $(X,Y,Z)$.
\end{property}

Because the coin is fair and independent, the selected loss $BX+(1-B)Z$ has distribution $\frac12\mu_X+\frac12\mu_Z$. Thus, Property (FI) is an independence condition for probabilistic mixtures, similar to the sure-thing principle of \citet{Savage1954}. A related condition is betweenness: if two losses have the same risk evaluation, then every probabilistic mixture of them has that evaluation. See \cite{Bellini2026} for a recent discussion and the connection with expectiles. For law-invariant risk measures, betweenness is the convex-level-sets property (CxLS). Under additional regularity conditions, CxLS is related to generalized shortfall risk measures \citep{DelbaenBelliniBignozziZiegel2016}.

In the proof of the following proposition, we show that a standard OCE satisfying Property (FI) also satisfies betweenness. Therefore, we apply the entropic classification \citet[Theorem~6.1]{Embrechts2021}. That result requires mixture Concavity, betweenness, and continuity at constant distributions. The preceding axioms provide the first condition, Property (FI) implies the second, and we verify the required continuity in the proof.

\begin{proposition}\label{prop:entropic}
A risk measure $\rho:L^\infty\to\R$ satisfies Axioms~(SRA), (FC), (N), and (M) and Property~(FI) if and only if there exists $\gamma\ge0$ such that
$$
\rho(X)=\rho_\gamma(X),\qquad X\in L^\infty,
$$
where $\rho_\gamma$ is the Entropic Risk Measure in \eqref{eq:entropic} for $\gamma>0$ and $\rho_0=\E$.
\end{proposition}

Therefore, Proposition~\ref{prop:entropic} gives an alternative axiomatization of the Entropic Risk Measures.

\section{Conclusion}\label{sec:conclusion}

We have shown that Strong Risk Aversion, Fair-Coin Compensation, and Normalization characterize OCEs, that is, risk measures of the form \eqref{eq:oce} with a convex loss function. Adding Monotonicity axiomatizes the standard OCEs, and Monotonicity requires the loss function to be nondecreasing. 

Economically, Fair-Coin Compensation is based on assigning two losses to two participants by an independent fair coin. Concavity in probabilistic mixtures implies that the random assignment creates a nonnegative fairness premium, and Fair-Coin Compensation requires that transfers on the coin can reduce this premium to zero.

Theorem~\ref{thm:main} characterizes the entire class but does not pick a specific risk measure. Section~\ref{sec:selecting} shows that additional properties select the most common members of the class: Equal-Mean Randomization selects the Mean--Variance risk measures, Positive Homogeneity and Prudence select Expected Shortfall, and Fair-Coin Independence selects the Entropic Risk Measures.

\section*{Acknowledgements}

CBW acknowledges financial support from the Natural Sciences and Engineering Research Council of Canada (RGPIN-2025-06879).

\section*{Use of artificial intelligence tools}

The authors used ChatGPT and Claude during manuscript preparation. The tools helped with brainstorming, language editing, and proof validation. In particular, ChatGPT suggested a mathematical version of Axiom~(FC) (combined with different versions of Axiom~(SRA)), but these versions were difficult to interpret or justify economically, and the authors reworked them into axioms that are more natural and easier to interpret in decision theory. The authors wrote and verified the proofs and take full responsibility for the final content.

\bibliographystyle{apalike}
\bibliography{references}

\newpage
\appendix
\renewcommand{\thelemma}{\Alph{section}.\arabic{lemma}}
\renewcommand{\thedefinition}{\Alph{section}.\arabic{definition}}
\renewcommand{\theequation}{\Alph{section}.\arabic{equation}}
\renewcommand{\theHlemma}{\Alph{section}.\arabic{lemma}}
\renewcommand{\theHdefinition}{\Alph{section}.\arabic{definition}}
\renewcommand{\theHequation}{\Alph{section}.\arabic{equation}}

\setcounter{equation}{0}
\setcounter{lemma}{0}
\setcounter{definition}{0}

\begin{center}
    \Large Appendices: Omitted proofs and additional results
\end{center}

\section{Proofs for Section  \ref{sec:axioms}}\label{app:A}

We first fix the notation that we use only in the appendices. We denote by $\mathcal D^1$ the set of Borel probability distributions on $\R$ with finite first absolute moment. We write
$$L^1_{\mathrm m}=\{Q\in L^1:\E[Q]=1\}~~\text{and}~~\mathcal D^1_{\mathrm m}=\left\{\mu\in\mathcal D^1:\int_{\R}z\,\mu(\mathrm dz)=1\right\},$$
where the subscript $\mathrm m$ imposes mean one. A random variable is simple if it takes finitely many values, and a finite $\sigma$-algebra is a $\sigma$-algebra generated by a finite partition of $\Omega$. For a sub-$\sigma$-algebra $\mathcal G\subseteq\cF$, we write $L^\infty(\mathcal G)$ for the set of essentially bounded $\mathcal G$-measurable random variables. The support of a distribution $\mu$ on $\R$ is
$$\operatorname{supp}\mu=\{z\in\R:\mu((z-\varepsilon,z+\varepsilon))>0\text{ for every }\varepsilon>0\},$$
and $\mu$ is finitely supported if its support is finite, equivalently if $\mu(F)=1$ for some finite $F\subseteq\R$. We write $\sigma(L^1,L^\infty)$ for the coarsest topology on $L^1$ for which $Q\mapsto\E[QX]$ is continuous for every $X\in L^\infty$. For a function $g$ with values in $\R\cup\{+\infty\}$, we write $\operatorname{dom}g=\{x:g(x)<\infty\}$ for its domain. For a convex function $g:\R\to\R\cup\{+\infty\}$ and $x\in\operatorname{dom}g$, the subdifferential $\partial g(x)$ is the set of $\kappa\in\R$ such that $g(z)\ge g(x)+\kappa(z-x)$ for every $z\in\R$. For $\ell\in\cL$, we write $\ell^*(z)=\sup_{x\in\R}\{xz-\ell(x)\}$ for the convex conjugate of $\ell$. It is convex and lower semicontinuous; it is nonnegative because $\ell(0)=0$, and it vanishes at $z=1$ because $\ell(x)\ge x$.

We use the fact that a finite convex function on $\R^n$ is continuous and has a subgradient at every point, and a proper, lower semicontinuous, convex function coincides with its biconjugate \citep[Corollary~10.1.1 and Theorems~12.2 and~23.4]{Rockafellar1970}.

\begin{proof}[Proof of Proposition~\ref{prop:properties}]
Law invariance follows from Axiom~(SRA). Fix $X\in L^\infty$ and $c\in\R$. By Assumption~\ref{ass:atomless}, we can find $X'\eqd X$ and a fair coin $B$ that is independent of $X'$. Applying Axiom~(FC) to $(X'+c,-c)$ and to $(X',0)$ gives
$$
\begin{aligned}
\frac{1}{2}\rho(X'+c)+\frac{1}{2}\rho(-c)
&=\inf_{a\in\R}\rho\left(B(X'+c+a)+(1-B)(-c-a)\right)\\
&=\inf_{b\in\R}\rho\left(B(X'+b)+(1-B)(-b)\right)\\
&=\frac{1}{2}\rho(X')+\frac{1}{2}\rho(0),
\end{aligned}
$$
where the second equality uses the change of variable $b=a+c$. By law invariance and Axiom~(N),
$$\rho(X+c)=\rho(X)+\rho(0)-\rho(-c)=\rho(X)+c.$$
Hence $\rho$ is cash additive.

We next prove convexity in outcome mixtures. Let $X,Y\in L^\infty$. By Assumption~\ref{ass:atomless}, we can find $(X',Y')\eqd(X,Y)$ and a fair coin $B$ that is independent of $(X',Y')$. For $a\in\R$, define the random variable $W_a=B(X'+a)+(1-B)(Y'-a)$. Since
$$\E[W_a\mid X',Y']=\frac{1}{2}X'+\frac{1}{2}Y',$$
conditional Jensen's inequality gives $(1/2)X'+(1/2)Y'\lecx W_a$ for every $a\in\R$. Law invariance and Axioms~(SRA) and (FC) therefore give
$$
\rho\left(\frac{1}{2}X+\frac{1}{2}Y\right)
=\rho\left(\frac{1}{2}X'+\frac{1}{2}Y'\right)\le\inf_{a\in\R}\rho(W_a)
=\frac{1}{2}\rho(X')+\frac{1}{2}\rho(Y')=\frac{1}{2}\rho(X)+\frac{1}{2}\rho(Y).
$$
This midpoint inequality implies, by induction on $n$,
\begin{equation}\label{eq:dyadic}
\rho(tX+(1-t)Y)\le t\rho(X)+(1-t)\rho(Y)
\end{equation}
for every dyadic weight $t=k/2^n$, where $n\ge0$ and $k\in\{0,\dots,2^n\}$.

To extend \eqref{eq:dyadic} to every $t\in[0,1]$, we first establish continuity with respect to $\|\cdot\|_\infty$. By Assumption~\ref{ass:atomless}, fix a random variable $S$ with $\P(S=-1)=\P(S=1)=1/2$.
Let $K>0$ and $Z\in L^\infty$ satisfy $\|Z\|_\infty\le K$, and set $\bar Z=Z-\E[Z]$. We have $\E[\bar Z]=0$ and $|\bar Z|\le2K$. For $z\in[-2K,2K]$, write
$$z=\frac{2K-z}{4K}(-2K)+\frac{2K+z}{4K}(2K).$$
Hence every convex $\phi:\R\to\R$ satisfies
$$\phi(z)\le\frac{2K-z}{4K}\phi(-2K)+\frac{2K+z}{4K}\phi(2K).$$
Taking expectations gives
$$\E[\phi(\bar Z)]\le\frac{2K-\E[\bar Z]}{4K}\phi(-2K) +\frac{2K+\E[\bar Z]}{4K}\phi(2K)=\frac{1}{2}\phi(-2K)+\frac{1}{2}\phi(2K)=\E[\phi(2KS)],$$
Thus $\bar Z\lecx2KS$, and cash additivity and Axiom~(SRA) imply
\begin{equation}\label{eq:ballbound}
\rho(Z)=\E[Z]+\rho(\bar Z)\le K+\rho(2KS).
\end{equation}

Fix $X\in L^\infty$ and write $g(H)=\rho(X+H)-\rho(X)$. The function $g$ inherits midpoint convexity from $\rho$ and satisfies $g(0)=0$. Put $K=\|X\|_\infty+1$ and $C=K+\rho(2KS)-\rho(X)$. By \eqref{eq:ballbound}, we have $C\ge0$. If $\|H\|_\infty\le1$, then $\|X+H\|_\infty\le K$, so, by \eqref{eq:ballbound}, we have that
$$g(H)=\rho(X+H)-\rho(X)\le K+\rho(2KS)-\rho(X)=C.$$
Additionally, for any $H\in L^\infty$, midpoint convexity gives
$$
\begin{aligned}
g(H/2)=\rho\left(\frac{1}{2}(X+H)+\frac{1}{2}X\right)-\rho(X)\le\frac{1}{2}\rho(X+H)+\frac{1}{2}\rho(X)-\rho(X)
=\frac{1}{2}g(H).
\end{aligned}
$$
Applying this inequality successively gives $g(2^{-n}H)\le2^{-n}g(H).$ Therefore, if $\|H\|_\infty\le1$, then
$$g(2^{-n}H)\le2^{-n}g(H)\le2^{-n}C.$$
Also, if $\|H\|_\infty\le1$, midpoint convexity gives
$$0=g(0)\le\frac{1}{2}g(2^{-n}H)+\frac{1}{2}g(-2^{-n}H)\implies g(-2^{-n}H)\ge-g(2^{-n}H)\ge-2^{-n}C.$$
Therefore, for any $H\in L^\infty$ with $\|H\|_\infty\le2^{-n}$, we have
$$|\rho(X+H)-\rho(X)|\le2^{-n}C.$$
Therefore, $\rho$ is continuous with respect to $\|\cdot\|_\infty$.

Finally, let $t\in[0,1]$ and choose a sequence $(t_n)_{n\in\mathbb{N}}$ of dyadic numbers converging to $t$. Since
$$\lim_{n\to\infty}\|(t_nX+(1-t_n)Y)-(tX+(1-t)Y)\|_\infty =\lim_{n\to\infty}|t_n-t|\,\|X-Y\|_\infty=0,$$
norm continuity and \eqref{eq:dyadic} give
$$\begin{aligned}
\rho(tX+(1-t)Y)=\lim_{n\to\infty}\rho(t_nX+(1-t_n)Y)&\le\lim_{n\to\infty}\{t_n\rho(X)+(1-t_n)\rho(Y)\}\\&=t\rho(X)+(1-t)\rho(Y).
\end{aligned}$$
Hence $\rho$ is convex in outcome mixtures.

We now prove concavity in probabilistic mixtures. Since $\rho$ is law invariant, we can write it as a functional $\widehat\rho$ on distributions, with $\widehat\rho(\mu_X)=\rho(X)$. Fix $\mu,\nu\in\mathcal D^\infty$. By Assumption~\ref{ass:atomless}, we can choose $X,Y$ with distributions $\mu,\nu$ and a uniform random variable $U$ on $[0,1]$ that is independent of $(X,Y)$. Then $B=\mathbbm{1}_{\{U\le1/2\}}$ is an independent fair coin. Axiom~(FC) gives
$$
\begin{aligned}
\widehat\rho\left(\frac{1}{2}\mu+\frac{1}{2}\nu\right)=\rho\left(BX+(1-B)Y\right)&\ge\inf_{a\in\R}\rho\left(B(X+a)+(1-B)(Y-a)\right)\\
&=\frac{1}{2}\widehat\rho(\mu)+\frac{1}{2}\widehat\rho(\nu).
\end{aligned}
$$
This midpoint inequality implies, by induction on $n$,
\begin{equation}\label{eq:dyadic-laws}
\widehat\rho(q\mu+(1-q)\nu)\ge q\widehat\rho(\mu)+(1-q)\widehat\rho(\nu)
\end{equation}
for every dyadic weight $q=k/2^n$, where $n\ge0$ and $k\in\{0,\dots,2^n\}$. Fix $t\in(0,1)$ and choose a sequence $(q_n)_{n\in\mathbb{N}}$ of dyadic numbers decreasing to $t$, with $t<q_n<1$. Set
$$B_t=\mathbbm{1}_{\{U\le t\}},~~ B_n=\mathbbm{1}_{\{U\le q_n\}},~~ Z=B_tX+(1-B_t)Y.$$
The distribution of $Z$ is $t\mu+(1-t)\nu$. Note that
$$\begin{aligned}
\P(B_t=1\mid B_n=1)
=\frac{\P(U\le t,\,U\le q_n)}{\P(U\le q_n)}
=\frac{t}{q_n}
\end{aligned}$$
On $\{B_n=0\}$, we have $U>q_n>t$, so $B_t=0$. Hence $\E[B_t\mid X,Y,B_n]=tB_n/q_n.$ Define
$$X_n=\frac{t}{q_n}X+\left(1-\frac{t}{q_n}\right)Y.$$
Writing $Z=Y+B_t(X-Y)$ and taking conditional expectations, we obtain
$$
\begin{aligned}
\E[Z\mid X,Y,B_n]=Y+(X-Y)\E[B_t\mid X,Y,B_n]&=Y+\frac{t}{q_n}B_n(X-Y)\\&=B_nX_n+(1-B_n)Y=:W_n.
\end{aligned}
$$
Conditional Jensen's inequality yields $W_n\lecx Z$, so Axiom~(SRA) gives $\rho(W_n)\le\rho(Z)$. Since $B_n$ is independent of $(X_n,Y)$ and $q_n$ is dyadic, \eqref{eq:dyadic-laws} implies
$$
\begin{aligned}
\widehat\rho(t\mu+(1-t)\nu)=\rho(Z)\ge\rho(W_n)\ge q_n\rho(X_n)+(1-q_n)\rho(Y).
\end{aligned}
$$
Moreover,
$$\lim_{n\to\infty}\|X_n-X\|_\infty =\lim_{n\to\infty}\left(1-\frac{t}{q_n}\right)\|Y-X\|_\infty=0.$$
Norm continuity gives $\rho(X_n)\to\rho(X)$. Passing to the limit in the preceding inequality, we obtain
$$\widehat\rho(t\mu+(1-t)\nu)\ge t\widehat\rho(\mu)+(1-t)\widehat\rho(\nu).$$
Therefore $\rho$ is concave in probabilistic mixtures.

Finally, we have shown that $\rho$ is law invariant, convex, and continuous with respect to $\|\cdot\|_\infty$. Since the probability space is atomless by Assumption~\ref{ass:atomless}, \citet[Theorem~1]{Liebrich2025} shows that $\rho$ is weakly lower semicontinuous.
\end{proof}

\begin{proof}[Proof of Lemma \ref{lem:ref}]
Let $C=(X+Y)/2$, $D=(X-Y)/2$, and $S=2B-1$, where
$B$ is the fair coin used in the assignment, independent
of $(X,Y)$. Set $K=\max\{\|X\|_\infty,\|Y\|_\infty\}$.
Then $|D|\le K$ and
\begin{equation}\label{eq:justneed}
    W_1(a)=C+S(D+a).
\end{equation}
Since $W_1(a)\eqd W_2(a)$ and Axiom~(SRA) implies
law invariance, we have $R(a)=2\rho(W_1(a))$.

For any convex function $\phi:\R\to\R$ and $c\in\R$,
the function
$$
h_c(t)=\frac12\phi(c+t)+\frac12\phi(c-t)
$$
is convex and even, and hence nondecreasing in $|t|$.
By independence,
$$
\E[\phi(W_1(a))]=\E[h_C(D+a)].
$$
If $a>K$, then $0\le D+K\le D+a$, so
$
\E[\phi(W_1(K))]\le\E[\phi(W_1(a))].
$
As this holds for every convex function $\phi$, we obtain
$W_1(K)\lecx W_1(a)$. Similarly, if $a<-K$, then
$D+a\le D-K\le0$, which yields
$W_1(-K)\lecx W_1(a)$. Axiom~(SRA) therefore gives
$R(K)\le R(a)$ for $a>K$ and $R(-K)\le R(a)$ for
$a<-K$. Consequently,
$$
\inf_{a\in\R}R(a)=\inf_{a\in[-K,K]}R(a).
$$

Finally, let $(a_n)_{n\in\mathbb{N}}$ be a minimizing
sequence in $[-K,K]$.
By compactness, there exists a subsequence $(a_{n_k})_{k\in \mathbb{N}}$
converging to some $a^*\in[-K,K]$. By \eqref{eq:justneed}, we have that
$$
\|W_1(a_{n_k})-W_1(a^*)\|_\infty
=\|S(a_{n_k}-a^*)\|_\infty= |a_{n_k}-a^*|.
$$
Thus $W_1(a_{n_k})$ converges weakly to $W_1(a^*)$.
By weak lower semicontinuity,
$$
R(a^*)=2\rho(W_1(a^*))\le 2\liminf_{k\to\infty}\rho(W_1(a_{n_k}))=\inf_{a\in\R}R(a).
$$
Therefore, the proof is complete.
\end{proof}

\begin{proof}[Proof of Proposition~\ref{prop:VaR-FC}]
Fix $p\in(0,1)$. Let $X=0$, and let $Y\in\{0,1\}$ satisfy
$$
\P(Y=0)=p^2,\qquad \P(Y=1)=1-p^2.
$$
Since $p^2<p$, we have $\VaR_p(X)=0$ and $\VaR_p(Y)=1$. Let $B$ be a fair coin independent of $Y$. Then
$$
\P((1-B)Y=0)
=\frac{1+p^2}{2}
=p+\frac{(1-p)^2}{2}
>p,
$$
so $\VaR_p((1-B)Y)=0$. Consequently, taking the transfer $a=0$ in Axiom~(FC) gives
$$
\inf_{a\in\R}
\VaR_p\bigl(B(X+a)+(1-B)(Y-a)\bigr)
\le
\VaR_p\bigl(BX+(1-B)Y\bigr)
=0
<
\frac12
=
\frac{\VaR_p(X)+\VaR_p(Y)}{2}.
$$
Thus the equality required by Axiom~(FC) fails, proving the proposition.
\end{proof}

The proofs of Theorems~\ref{thm:main} and~\ref{thm:oce} use the dual representation of risk measures through penalties on densities, following \citet{Follmer2002}.

\begin{definition}[Minimal penalty]\label{def:minimal-penalty}
For a functional $\rho:L^\infty\to\R$, its minimal penalty is the functional $\alpha_\rho:L^1\to\R\cup\{+\infty\}$ defined by
$$\alpha_\rho(Q)=\sup_{X\in L^\infty}\left\{\E[QX]-\rho(X)\right\},~~Q\in L^1.$$
\end{definition}

For later reference, we collect standard properties of minimal penalties using the results of \citet{Bellini2021}. The result numbers below refer to the arXiv version.

\begin{lemma}\label{lem:penalty-working}
Let $\rho: L^\infty \to \R$ be law invariant, cash additive, convex in outcome mixtures, and lower semicontinuous with respect to $\sigma(L^\infty, L^1)$, with $\rho(0)=0$. Then its minimal penalty has the following properties.
\begin{enumerate}
\item[(i)] $\alpha_\rho\ge0$, $\alpha_\rho(1)=0$, and $\operatorname{dom}\alpha_\rho\subseteq L^1_{\mathrm m}$.
\item[(ii)] $\alpha_\rho$ is convex and lower semicontinuous with respect to the $L^1$ norm.
\item[(iii)] If $Q\in L^1$ is measurable with respect to a sub-$\sigma$-algebra $\mathcal G\subseteq\cF$, then
$$\alpha_\rho(Q)=\sup_{X\in L^\infty(\mathcal G)}\left\{\E[QX]-\rho(X)\right\}.$$
\item[(iv)] $\alpha_\rho$ is law invariant: $\alpha_\rho(Q)=\alpha_\rho(Q')$ whenever $Q,Q'\in L^1$ satisfy $Q\eqd Q'$.
\item[(v)] For every $Q\in L^1$ and every sub-$\sigma$-algebra $\mathcal G\subseteq\cF$, $\alpha_\rho(\E[Q\mid\mathcal G])\le\alpha_\rho(Q).$
\item[(vi)] For every $X\in L^\infty$, $\rho(X)=\sup_{Q\in L^1_{\mathrm m}}\left\{\E[QX]-\alpha_\rho(Q)\right\}.$
\end{enumerate}
\end{lemma}

\begin{proof}
By \citet[Proposition~5.6]{Bellini2021}, $\rho$ is dilatation monotone, so
$$\rho(\E[X\mid\mathcal G])\le\rho(X)$$
for every $X\in L^\infty$ and every sub-$\sigma$-algebra $\mathcal G\subseteq\cF$.

(i) Normalization and cash additivity give $\rho(c)=c$ for every $c\in\R$. Taking $X=0$ in the definition gives $\alpha_\rho(Q)\ge0$, while taking $X=c$ gives
$$\alpha_\rho(Q)\ge c\bigl(\E[Q]-1\bigr)~~c\in\R.$$
Thus $\alpha_\rho(Q)=+\infty$ if $\E[Q]\ne1$. Applying dilatation monotonicity with the trivial $\sigma$-algebra gives $\rho(X)\ge\rho(\E[X])=\E[X]$, so $\alpha_\rho(1)=0$.

(ii) By its definition, $\alpha_\rho$ is a supremum of $\sigma(L^1,L^\infty)$-continuous affine functions. It is therefore convex and $\sigma(L^1, L^\infty)$-lower semicontinuous, and hence $L^1$-norm lower semicontinuous.

(iii) If $Q$ is $\mathcal G$-measurable, then dilatation monotonicity gives
$$
\begin{aligned}
\E[QX]-\rho(X)
&=\E[Q\E[X\mid\mathcal G]]-\rho(X)\le\E[Q\E[X\mid\mathcal G]]-\rho(\E[X\mid\mathcal G]).
\end{aligned}
$$
Thus restricting the supremum to $L^\infty(\mathcal G)$ leaves its value unchanged.

(iv) Since $\alpha_\rho$ is the conjugate of $\rho$ for the dual pair $(L^\infty,L^1)$, this follows from \citet[Corollary~5.2(i)]{Bellini2021}.

(v) By (i) and (iv), $\alpha_\rho$ is proper and law invariant. Its convexity and $\sigma(L^1, L^\infty)$-lower semicontinuity were established above. Applying \citet[Proposition~5.6]{Bellini2021} with the dual pair $(L^1,L^\infty)$ gives the result.

(vi) This follows from Fenchel--Moreau duality.
\end{proof}

\begin{lemma}\label{lem:oce-penalty-working}
Let $\ell\in\mathcal L$. Then $\rho_\ell:L^\infty\to\R$ satisfies the hypotheses of Lemma~\ref{lem:penalty-working}, and
$$\alpha_{\rho_\ell}(Q)=\E[\ell^*(Q)],~~Q\in L^1_{\mathrm m}.$$
\end{lemma}

\begin{proof}
Law invariance and $\rho_\ell(0)=0$ follow directly from the definition. Cash additivity and convexity in outcome mixtures follow from the arguments in the proof of \citet[Theorem~2.1(a), (d)]{BenTal2007}, with $u(x)=-\ell(-x)$.

Since $\operatorname{dom}\ell$ is an unbounded interval containing $0$, it contains either $[0,\infty)$ or $(-\infty,0]$. Fix $K>0$ and $X\in L^\infty$ with $\|X\|_\infty\le K$.
If $[0,\infty)\subseteq\operatorname{dom}\ell$, convexity of $\ell$ gives
$$\E[X]\le\rho_\ell(X)\le-K+\E[\ell(X+K)]\le-K+\max\{\ell(0),\ell(2K)\}<\infty.$$
The lower bound follows from $\ell(x)\ge x$. If $(-\infty,0]\subseteq\operatorname{dom}\ell$, then a similar argument gives $\rho_\ell(X)\le K+\max\{\ell(-2K),\ell(0)\}<\infty$. Thus $\rho_\ell$ is locally bounded above. By \citet[Proposition 4.1.4]{BV10}, convexity and local boundedness implies continuity with respect to $\|\cdot\|_{\infty}$. 
By \citet[Theorem~1]{Liebrich2025}, $\rho_\ell$ is lower semicontinuous with respect to $\sigma(L^\infty,L^1)$.

Finally, fix $Q\in L^1_{\mathrm m}$. The definitions give
$$
\begin{aligned}
\alpha_{\rho_\ell}(Q)
&=\sup_{X\in L^\infty}\sup_{m\in\R}\left\{\E[QX]-m-\E[\ell(X-m)]\right\}\\
&=\sup_{Y\in L^\infty}\left\{\E[QY]-\E[\ell(Y)]\right\}=\E[\ell^*(Q)],
\end{aligned}
$$
where the last equality follows from \citet[Corollary to Theorem~2]{Rockafellar1968}.
\end{proof}

If $\rho:L^\infty\to\R$ satisfies the hypotheses of Lemma~\ref{lem:penalty-working}, then, by part (iv) and Assumption~\ref{ass:atomless}, its minimal penalty induces a well-defined functional
$$\widehat\alpha_\rho:\mathcal D^1_{\mathrm m}\to[0,+\infty],~~ \widehat\alpha_\rho(\mu_Q)=\alpha_\rho(Q),~~Q\in L^1_{\mathrm m}.$$

\begin{lemma}\label{lem:penalty-mixtures}
Let $\rho: L^\infty \to \R$ satisfy Axioms~(SRA), (FC), and (N), and let $\mu,\nu\in\mathcal D^1_{\mathrm m}$. Then, 
\begin{equation}\label{eq:penalty-mixtures}
\widehat\alpha_\rho(t\mu+(1-t)\nu)=t\widehat\alpha_\rho(\mu)+(1-t)\widehat\alpha_\rho(\nu),~~t\in[0,1].
\end{equation}
\end{lemma}

\begin{proof}
By Proposition~\ref{prop:properties} and Axiom~(N), Lemma~\ref{lem:penalty-working} applies to $\rho$. By Assumption~\ref{ass:atomless}, choose $Q_1,Q_2\in L^1_{\mathrm m}$ with distributions $\mu,\nu$ and a random variable $U$ uniform on $[0,1]$ and independent of $(Q_1,Q_2)$. Set $$B=\mathbbm{1}_{\{U\le1/2\}}~~\text{and}~~Q=BQ_1+(1-B)Q_2,$$ whose distribution is $(1/2)\mu+(1/2)\nu$. Let $\mathcal G=\sigma(B,BQ_1,(1-B)Q_2)$. Then $Q$ is $\mathcal G$-measurable.

For any $Z\in L^\infty(\mathcal G)$, the Doob--Dynkin lemma gives $Z=g(B,BQ_1,(1-B)Q_2)$ for a bounded Borel function $g$. Taking $X=g(1,Q_1,0)$ and $Y=g(0,0,Q_2)$ gives $Z=BX+(1-B)Y$, with $X\in L^\infty(\sigma(Q_1))$ and $Y\in L^\infty(\sigma(Q_2))$. Conversely, every loss of this form is $\mathcal G$-measurable. 

Independence gives
$$\E[Q(BX+(1-B)Y)]=\frac12\E[Q_1X]+\frac12\E[Q_2Y]=:\Phi(X,Y).$$
Let $\mathcal H=\{(X,Y):X\in L^\infty(\sigma(Q_1)),\,Y\in L^\infty(\sigma(Q_2))\}$. Since $\E[Q_1]=\E[Q_2]=1$, we have $\Phi(X+a,Y-a)=\Phi(X,Y)$ for every $a\in\R$. Lemma~\ref{lem:penalty-working}(iii) and Axiom~(FC) therefore give
$$
\begin{aligned}
\alpha_\rho(Q)
&=\sup_{(X,Y)\in\mathcal H}\left\{\Phi(X,Y)-\rho\left(BX+(1-B)Y\right)\right\}\\
&=\sup_{(X,Y)\in\mathcal H}\sup_{a\in\R}\left\{\Phi(X,Y)-\rho\left(B(X+a)+(1-B)(Y-a)\right)\right\}\\
&=\sup_{(X,Y)\in\mathcal H}\left\{\Phi(X,Y)-\inf_{a\in\R}\rho\left(B(X+a)+(1-B)(Y-a)\right)\right\}\\
&=\sup_{(X,Y)\in\mathcal H}\left\{\frac12\bigl(\E[Q_1X]-\rho(X)\bigr)+\frac12\bigl(\E[Q_2Y]-\rho(Y)\bigr)\right\}\\
&=\frac12\alpha_\rho(Q_1)+\frac12\alpha_\rho(Q_2).
\end{aligned}
$$
The last equality uses Lemma~\ref{lem:penalty-working}(iii) with $\sigma(Q_1)$ and $\sigma(Q_2)$. Thus \eqref{eq:penalty-mixtures} holds for $t=1/2$.

This midpoint identity implies, by induction on $n$, that \eqref{eq:penalty-mixtures} holds for every dyadic weight $t=k/2^n$, where $n\ge0$ and $k\in\{0,\dots,2^n\}$. Fix $t\in(0,1)$ and choose a sequence $(q_n)_{n\in\mathbb{N}}$ of dyadic numbers decreasing to $t$, with $t<q_n<1$. For $s\in[0,1]$, set
$$Q_s=\mathbbm{1}_{\{U\le s\}}Q_1+\mathbbm{1}_{\{U>s\}}Q_2.$$
Since
$$\lim_{n\to\infty}\|Q_{q_n}-Q_t\|_1=\lim_{n\to\infty}(q_n-t)\,\E|Q_1-Q_2|=0,$$
Lemma~\ref{lem:penalty-working}(ii) and the dyadic identity give
$$\widehat\alpha_\rho(t\mu+(1-t)\nu)\le\liminf_{n\to\infty}\alpha_\rho(Q_{q_n})=t\widehat\alpha_\rho(\mu)+(1-t)\widehat\alpha_\rho(\nu).$$
If $\widehat\alpha_\rho(\mu)$ and $\widehat\alpha_\rho(\nu)$ are finite, the preceding inequality makes $\widehat\alpha_\rho(t\mu+(1-t)\nu)$ finite for every $t\in[0,1]$. The fair-coin identity also gives
$$\widehat\alpha_\rho(t\mu+(1-t)\nu)+\widehat\alpha_\rho((1-t)\mu+t\nu)=\widehat\alpha_\rho(\mu)+\widehat\alpha_\rho(\nu).$$
Therefore, for $t\in(0,1)$, we have
$$
\begin{aligned}
\widehat\alpha_\rho(t\mu+(1-t)\nu)
&=\widehat\alpha_\rho(\mu)+\widehat\alpha_\rho(\nu)-\widehat\alpha_\rho((1-t)\mu+t\nu)\\
&\ge\widehat\alpha_\rho(\mu)+\widehat\alpha_\rho(\nu)-\bigl((1-t)\widehat\alpha_\rho(\mu)+t\widehat\alpha_\rho(\nu)\bigr)\\
&=t\widehat\alpha_\rho(\mu)+(1-t)\widehat\alpha_\rho(\nu).
\end{aligned}
$$
Therefore \eqref{eq:penalty-mixtures} holds when both endpoint penalties are finite.

If $\widehat\alpha_\rho(\mu)=+\infty$, choose a dyadic $d\in(0,t)$ and put $s=(t-d)/(1-d)\in[0,1]$. This choice gives $t=d+(1-d)s$, and hence
$$t\mu+(1-t)\nu=d\mu+(1-d)\bigl(s\mu+(1-s)\nu\bigr).$$
We apply the dyadic identity of \eqref{eq:penalty-mixtures} to $\mu$ and $s\mu+(1-s)\nu$ and obtain
$$\widehat\alpha_\rho(t\mu+(1-t)\nu)=d\widehat\alpha_\rho(\mu)+(1-d)\widehat\alpha_\rho(s\mu+(1-s)\nu)=+\infty.$$
The case $\widehat\alpha_\rho(\nu)=+\infty$ follows by interchanging $\mu$ and $\nu$. 
\end{proof}

\begin{proof}[Proof of Theorem~\ref{thm:main}]
Let $\rho$ satisfy Axioms~(SRA), (FC), and (N). Then $\rho$ satisfies the hypotheses of Lemma~\ref{lem:penalty-working}, and its minimal penalty $\alpha_\rho$ has all the properties stated there. Moreover, $\widehat\alpha_\rho$ satisfies the mixture identity in Lemma~\ref{lem:penalty-mixtures}.

Let $A$ be the set of points $z\in\R$ for which some finitely supported $\mu\in\mathcal D^1_{\mathrm m}$ satisfies $\mu(\{z\})>0$ and $\widehat\alpha_\rho(\mu)<\infty$. By Lemma~\ref{lem:penalty-working}(i), $\widehat\alpha_\rho(\delta_1)=\alpha_\rho(1)=0$, so $1\in A$.

We claim that every finitely supported $\mu\in\mathcal D^1_{\mathrm m}$ supported in $A$ has finite penalty. To see this, write $\operatorname{supp}\mu=\{z_1,\ldots,z_n\}$. For each $k=1,\ldots,n$, choose a finitely supported $\mu_k\in\mathcal D^1_{\mathrm m}$ with $\mu_k(\{z_k\})>0$ and $\widehat\alpha_\rho(\mu_k)<\infty$, and define
$$\bar\mu=\frac{1}{n}\sum_{k=1}^n\mu_k.$$
Lemma~\ref{lem:penalty-mixtures} gives $\widehat\alpha_\rho(\bar\mu)<\infty$, and $\bar\mu(\{z_k\})\ge(1/n)\mu_k(\{z_k\})>0$ for every $k=1,\ldots,n$. Choose $\varepsilon\in(0,1)$ so that $\bar\mu\ge\varepsilon\mu$, and put $\nu=(\bar\mu-\varepsilon\mu)/(1-\varepsilon)\in\mathcal D^1_{\mathrm m}$. Then
$$\widehat\alpha_\rho(\bar\mu)=\varepsilon\widehat\alpha_\rho(\mu)+(1-\varepsilon)\widehat\alpha_\rho(\nu)<\infty.$$
Since the penalties are nonnegative, $\widehat\alpha_\rho(\mu)<\infty$.

The set $A$ is an interval. To see this, let $u,v\in A$ and $\theta\in(0,1)$. By Assumption~\ref{ass:atomless}, choose independent simple $Q_u,Q_v\in L^1_{\mathrm m}$ with finite penalties and $\P(Q_u=u)>0$, $\P(Q_v=v)>0$. By Lemma~\ref{lem:penalty-working}(ii),
$$\alpha_\rho(\theta Q_u+(1-\theta)Q_v)\le\theta\alpha_\rho(Q_u)+(1-\theta)\alpha_\rho(Q_v)<\infty.$$
Independence ensures that $\theta Q_u+(1-\theta)Q_v$ takes the value $\theta u+(1-\theta)v$ with positive probability, so this value belongs to $A$.

We claim that there exists a convex, lower semicontinuous function $f:\R\to[0,+\infty]$ with $f(1)=0$ such that $\alpha_\rho(Q)=\E[f(Q)]$ for every simple $Q\in L^1_{\mathrm m}$. If $A=\{1\}$, set $f(1)=0$ and $f(z)=+\infty$ for $z\ne1$. This function is nonnegative, convex, and lower semicontinuous, and by the definition of $ A$, $\alpha_\rho(Q)=\E[f(Q)]$ for every simple $Q\in L^1_{\mathrm m}$.

Otherwise, some finitely supported distribution with finite penalty has an atom different from $1$, and hence has atoms below and above $1$. Fix $a,b\in A$ with $a<1<b$. For $z\in A\setminus\{1\}$, let $\tau_z$ be the probability distribution with mean $1$ supported on $\{z,b\}$ if $z<1$ and on $\{z,a\}$ if $z>1$. Put $\tau_1=\delta_1$ and $w_z=\tau_z(\{z\})$, so that $0<w_z<1$ for $z\ne1$ and $w_1=1$. Since these distributions have support in $A$, they have finite penalties. Define $f_0:A\to\R$ by $f_0(1)=0$ and
$$
f_0(z)=\begin{cases}
\displaystyle\frac{\widehat\alpha_\rho(\tau_z)}{w_z},&z<1,\\[6pt]
\displaystyle\frac{\widehat\alpha_\rho(\tau_z)-(1-w_z)f_0(a)}{w_z},&z>1.
\end{cases}
$$
Since $\tau_a=\tau_b$ and $1-w_b=w_a$, the definition gives $f_0(b)=0$. For $z\in A$ with $z<1$, the distribution $\tau_z$ puts mass $w_z$ at $z$ and mass $1-w_z$ at $b$. The equality $f_0(b)=0$ therefore gives
$$\int_{\R} f_0\,\mathrm d\tau_z=w_zf_0(z)+(1-w_z)f_0(b)=w_zf_0(z)=\widehat\alpha_\rho(\tau_z).$$
For $z\in A$ with $z>1$, the other atom is $a$, and the second branch of the definition gives
$$\int_{\R} f_0\,\mathrm d\tau_z=w_zf_0(z)+(1-w_z)f_0(a)=\widehat\alpha_\rho(\tau_z).$$
For $z=1$, the same identity holds because $f_0(1)=\widehat\alpha_\rho(\delta_1)=0$.

Let $\mu\in\mathcal D^1_{\mathrm m}$ have finite support $\{z_1,\ldots,z_K\}\subseteq A$, write $\tau=\tau_a=\tau_b$, and set
$$\beta=\left(\sum_{k=1}^K\frac{\mu(\{z_k\})}{w_{z_k}}\right)^{-1}~~\text{and}~~ \lambda_k=\frac{\beta\mu(\{z_k\})}{w_{z_k}},~~k=1,\ldots,K.$$
Then $0<\beta\le1$ and $\sum_{k=1}^K\lambda_k=1$. We claim that
$\sum_{k=1}^K\lambda_k\tau_{z_k}=\beta\mu+(1-\beta)\tau.$
To see this, use $\lambda_k w_{z_k}=\beta\mu(\{z_k\})$ to write
$$\gamma=\sum_{k=1}^K\lambda_k(\tau_{z_k}-w_{z_k}\delta_{z_k})=\sum_{k=1}^K\lambda_k\tau_{z_k}-\beta\mu.$$
For $z_k<1$, the measure $\tau_{z_k}-w_{z_k}\delta_{z_k}$ equals $(1-w_{z_k})\delta_b$; for $z_k>1$, it equals $(1-w_{z_k})\delta_a$; and for $z_k=1$, it is zero. Hence $\gamma$ is a nonnegative measure supported on $\{a,b\}$. Since $\mu$ and all the $\tau_{z_k}$ have total mass and mean $1$, $\gamma=(1-\beta)\tau$, which proves the claim. Using Lemma~\ref{lem:penalty-mixtures}, we obtain\begin{align*}
\beta\widehat\alpha_\rho(\mu)+(1-\beta)\int_{\R} f_0\,\mathrm d\tau
&=\beta\widehat\alpha_\rho(\mu)+(1-\beta)\widehat\alpha_\rho(\tau)
\\&=\sum_{k=1}^K\lambda_k\widehat\alpha_\rho(\tau_{z_k})
\\&=\sum_{k=1}^K\lambda_k\int_{\R} f_0\,\mathrm d\tau_{z_k}\\&=\int_{\R} f_0\,\mathrm d\bigl(\beta\mu+(1-\beta)\tau\bigr)=\beta\int_{\R} f_0\,\mathrm d\mu+(1-\beta)\int_{\R} f_0\,\mathrm d\tau.
\end{align*}
Therefore, $\widehat\alpha_\rho(\mu)=\int_{\R} f_0\,\mathrm d\mu.$

To prove convexity of $f_0$, fix $u,v\in A$, $\theta\in(0,1)$, and $r=\theta u+(1-\theta)v$. Choose $\eta\in(0,1)$ small enough that $c=(1-\eta r)/(1-\eta)\in A$. By Assumption~\ref{ass:atomless}, choose a partition $E_1,E_2,E_3$ with probabilities $\eta\theta$, $\eta(1-\theta)$, and $1-\eta$, respectively. Put $Q=u\mathbbm{1}_{E_1}+v\mathbbm{1}_{E_2}+c\mathbbm{1}_{E_3}$ and $\mathcal G=\sigma(E_1\cup E_2)$. Then $\E[Q]=1$ and $\E[Q\mid\mathcal G]=r\mathbbm{1}_{E_1\cup E_2}+c\mathbbm{1}_{E_3}$. 
Therefore, by Lemma~\ref{lem:penalty-working}(v), we have that
\begin{align*}
\eta f_0(r)+(1-\eta)f_0(c)
=\E[f_0(\E[Q\mid\mathcal G])]
&=\alpha_\rho(\E[Q\mid\mathcal G])\\
&\le\alpha_\rho(Q)\\
&=\E[f_0(Q)]=\eta\theta f_0(u)+\eta(1-\theta)f_0(v)+(1-\eta)f_0(c).
\end{align*}
Thus $f_0$ is convex.

Since $1$ is an interior point of $A$, by \citet[Theorem~23.4]{Rockafellar1970}, there exists $\kappa\in\partial f_0(1)$. Define
$$
f(z)=\begin{cases}
f_0(z)-\kappa(z-1),&z\in A,\\
+\infty,&z\notin A.
\end{cases}
$$
Then $f$ is convex, $f\ge0$, and $f(1)=0$. For every finitely supported $\mu\in\mathcal D^1_{\mathrm m}$ with $\operatorname{supp}\mu\subseteq A$,
$$\int_{\R}f\,\mathrm d\mu=\int_{\R} f_0\,\mathrm d\mu=\widehat\alpha_\rho(\mu).$$
If a finitely supported distribution in $\mathcal D^1_{\mathrm m}$ has an atom outside $A$, its penalty is infinite by the definition of $A$ and its integral with respect to $f$ is infinite by the definition of $f$. Hence, for all simple $Q\in L^1_{\mathrm m}$, we have
\begin{equation}\label{eq:penalty-integral-simple}
\alpha_\rho(Q)=\E[f(Q)].
\end{equation}

By classic results on convex functions, $f$ is continuous on the interior of $A$ and infinite outside the closure of $A$, so we only need to check lower semicontinuity at the finite endpoints. Let $a_0=\inf A> -\infty$, and let $(z_n)_{n\in\mathbb{N}}$ be any sequence in $A\cap(a_0,1)$ with $\lim_{n\to\infty}z_n=a_0$. Set $p_n=(b-1)/(b-z_n)$ and $p=(b-1)/(b-a_0)>0$. By Assumption~\ref{ass:atomless}, choose $U$ uniform on $[0,1]$ and define
$$
Q_n=z_n\mathbbm{1}_{\{U\le p_n\}}+b\mathbbm{1}_{\{U>p_n\}},~~
Q=a_0\mathbbm{1}_{\{U\le p\}}+b\mathbbm{1}_{\{U>p\}}.
$$
Then $\lim_{n\to\infty}\|Q_n-Q\|_1=0$. Lemma~\ref{lem:penalty-working}(ii) and \eqref{eq:penalty-integral-simple} give
$$pf(a_0)+(1-p)f(b)=\alpha_\rho(Q)\le\liminf_{n\to\infty}\alpha_\rho(Q_n) =p\liminf_{n\to\infty}f(z_n)+(1-p)f(b). $$
Since $f(b)$ is finite and $p>0$, this yields $f(a_0)\le\liminf_{n\to\infty}f(z_n)$. The same argument at a finite upper endpoint, using $a$ in place of $b$, proves that $f$ is lower semicontinuous.

We claim that \eqref{eq:penalty-integral-simple} holds for every $Q\in L^1_{\mathrm m}$. To see this, enumerate the rational numbers as $(r_k)_{k\in\mathbb{N}}$ and set
$$\mathcal G_n=\sigma\bigl(\{Q\le r_k\}:1\le k\le n\bigr),~~Q_n=\E[Q\mid\mathcal G_n].$$
Then $(\mathcal G_n)_{n\in\mathbb{N}}$ is an increasing sequence of finite sub-$\sigma$-algebras with $\sigma(\bigcup_{n\in\mathbb{N}}\mathcal G_n)=\sigma(Q)$. Each $Q_n$ is simple and belongs to $L^1_{\mathrm m}$. By the upward martingale convergence theorem of \citet[Theorem~4.6.8]{Durrett2019}, $\lim_{n\to\infty}\|Q_n-Q\|_1=0.$ By Lemma~\ref{lem:penalty-working}(v), we have
$$\alpha_\rho(Q_n)\le\alpha_\rho(Q_{n+1})\le\alpha_\rho(Q).$$
Thus $(\alpha_\rho(Q_n))_{n\in\mathbb{N}}$ is increasing, and Lemma~\ref{lem:penalty-working}(ii) gives
\begin{equation}\label{eq:penalty-condexp-limit}
\alpha_\rho(Q)\le\liminf_{n\to\infty}\alpha_\rho(Q_n)=\lim_{n\to\infty}\alpha_\rho(Q_n)\le\alpha_\rho(Q).
\end{equation}
 
Define $\ell=f^*$. By scalar Fenchel--Moreau duality, $\ell^*=f$. Moreover, $\ell$ is proper, convex, and lower semicontinuous. Since $f\ge0$ and $f(1)=0$,
$$\ell(0)=-\inf_{z\in\R}f(z)=0~~\text{and}~~\ell(x)=\sup_{z\in\R}\{xz-f(z)\}\ge x-f(1)=x,~~x\in\R.$$
We claim that $\operatorname{dom}\ell$ is unbounded. To see this, suppose instead that $\operatorname{dom}\ell\subseteq[-M,M]$ for some $M>0$. Since $\ell(x)\ge x$,
$$f(z)=\sup_{x\in\operatorname{dom}\ell}\{xz-\ell(x)\}\le M|z-1|,~~z\in\R.$$
By Assumption~\ref{ass:atomless}, choose a fair coin $B$ and set $Z=2B-1$. For $s>0$, let $Q_s=1+sZ\in L^1_{\mathrm m}$. Since $Q_s$ is simple, \eqref{eq:penalty-integral-simple} gives $\alpha_\rho(Q_s)\le Ms$. By the definition of $\alpha_\rho$,
$$\rho((M+1)Z)\ge\E[Q_s(M+1)Z]-\alpha_\rho(Q_s)\ge(M+1)s-Ms=s,~~s>0,$$
contradicting the fact that $\rho$ is real valued. Thus $\ell\in\mathcal L$.

By Lemma~\ref{lem:oce-penalty-working}, $\rho_\ell$ satisfies the hypotheses of Lemma~\ref{lem:penalty-working}, and $\alpha_{\rho_\ell}(Q)=\E[f(Q)]$ for every $Q\in L^1_{\mathrm m}$. Repeating the preceding convergence argument with $\alpha_{\rho_\ell}$ in place of $\alpha_\rho$ gives
$$\lim_{n\to\infty}\E[f(Q_n)]=\lim_{n\to\infty}\alpha_{\rho_\ell}(Q_n)=\alpha_{\rho_\ell}(Q)=\E[f(Q)].$$
Combining this with \eqref{eq:penalty-integral-simple} and \eqref{eq:penalty-condexp-limit} yields
\begin{equation}\label{eq:penalty-integral}
\alpha_\rho(Q)=\E[f(Q)]=\alpha_{\rho_\ell}(Q),~~Q\in L^1_{\mathrm m}.
\end{equation}
Lemma~\ref{lem:penalty-working}(vi), applied to $\rho$ and $\rho_\ell$, therefore gives
\begin{equation}\label{eq:dual-working}
\rho(X)=\sup_{Q\in L^1_{\mathrm m}}\{\E[QX]-\alpha_\rho(Q)\}=\rho_\ell(X),~~X\in L^\infty.
\end{equation}

Conversely, let $\ell\in\mathcal L$. By Lemma~\ref{lem:oce-penalty-working}, $\rho_\ell$ is cash additive and $\rho_\ell(0)=0$, so $\rho_\ell(c)=c$ for every $c\in\R$, proving Axiom~(N). If $X\lecx Y$, convexity of $\ell$ gives $\E[\ell(X-m)]\le\E[\ell(Y-m)]$ for every $m\in\R$. Taking infima over $m$ proves Axiom~(SRA).

To prove Axiom~(FC), fix $X,Y\in L^\infty$ and a fair coin $B$ independent of $(X,Y)$. Independence and the bijective change of variables $u=m-a$, $v=m+a$ give
$$
\begin{aligned}
\inf_{a\in\R}\rho_\ell\bigl(B(X+a)+(1-B)(Y-a)\bigr)
&=\inf_{a,m\in\R}\left\{m+\frac12\E[\ell(X+a-m)]+\frac12\E[\ell(Y-a-m)]\right\}\\
&=\inf_{u,v\in\R}\left\{\frac12\bigl(u+\E[\ell(X-u)]\bigr)+\frac12\bigl(v+\E[\ell(Y-v)]\bigr)\right\}\\
&=\frac12\inf_{u\in\R}\{u+\E[\ell(X-u)]\}+\frac12\inf_{v\in\R}\{v+\E[\ell(Y-v)]\}\\
&=\frac12\rho_\ell(X)+\frac12\rho_\ell(Y).
\end{aligned}
$$
Therefore $\rho_\ell$ satisfies Axiom~(FC).

For uniqueness, first suppose that $\ell'\in\mathcal L$ is equivalent to $\ell$, so that $\ell'(x)=\ell(x+c)-c$ for some $c\in\R$. We have
$$\rho_{\ell'}(X)=\inf_{m\in\R}\{m-c+\E[\ell(X-m+c)]\}=\inf_{r\in\R}\{r+\E[\ell(X-r)]\}=\rho_\ell(X),~~X\in L^\infty.$$

Conversely, suppose that $\rho=\rho_\ell=\rho_{\ell'}\ne\E$, and set $f=\ell^*$ and $g=(\ell')^*$. Both functions are nonnegative and satisfy $f(1)=g(1)=0$. By Lemma~\ref{lem:oce-penalty-working},
$$\E[f(Q)]=\alpha_\rho(Q)=\E[g(Q)],~~Q\in L^1_{\mathrm m}.$$
Choose $a_0<1<b_0$ such that $f(a_0),f(b_0)<\infty$; otherwise, we would have $\rho=\E$. For $a<1<b$, Assumption~\ref{ass:atomless} allows us to choose a random variable $Q_{a,b}$ with
$$\P(Q_{a,b}=a)=\frac{b-1}{b-a},~~\P(Q_{a,b}=b)=\frac{1-a}{b-a}.$$
Since $\E[Q_{a,b}]=1$, equality of the penalties gives
\begin{equation}\label{eq:penalty-two-point-uniqueness}
(b-1)f(a)+(1-a)f(b)=(b-1)g(a)+(1-a)g(b).
\end{equation}
Taking $a=a_0$ and $b=b_0$ shows that $g(a_0),g(b_0)<\infty$. Fixing $b=b_0$ in \eqref{eq:penalty-two-point-uniqueness} now shows that $f(a)<\infty$ if and only if $g(a)<\infty$ for every $a<1$. Similarly, fixing $a=a_0$ gives the same equivalence for every $b>1$. Thus $f$ and $g$ have the same domain, which we denote by $D$.

For $a,b\in D$ with $a<1<b$, all terms in \eqref{eq:penalty-two-point-uniqueness} are finite, and rearranging gives
$$\frac{f(a)-g(a)}{a-1}=\frac{f(b)-g(b)}{b-1}.$$
Therefore, these ratios equal a single constant; denote it by $c\in\R$. Since $f(1)=g(1)=0$ and both functions are infinite outside $D$, we obtain
$$g(z)=f(z)-c(z-1),~~z\in\R.$$
Scalar Fenchel--Moreau duality therefore gives
$$\ell'(x)=g^*(x)=\sup_{z\in\R}\{xz-f(z)+c(z-1)\}=\ell(x+c)-c,~~x\in\R.$$
Hence $\ell'$ is equivalent to $\ell$, proving uniqueness.
\end{proof}

\begin{proof}[Proof of Theorem~\ref{thm:oce}]
By Theorem~\ref{thm:main}, every standard OCE satisfies Axioms~(SRA), (FC), and (N). Since its loss function is nondecreasing, it also satisfies Axiom~(M).

Conversely, suppose that $\rho$ satisfies the four axioms. By Theorem~\ref{thm:main}, $\rho=\rho_\ell$ for some $\ell\in\mathcal L$. If $\rho=\E$, we may take $\ell(x)=x\in\mathcal L_+$. Suppose that $\rho\ne\E$, and set $f=\ell^*$. As shown in the uniqueness argument in the proof of Theorem~\ref{thm:main}, there exists $b_0>1$ such that $f(b_0)<\infty$.

We claim that $f(z)=+\infty$ for every $z<0$. To see this, first let $Q\in L^1_{\mathrm m}$ satisfy $\P(Q<0)>0$. By Axioms~(M) and (N), $\rho(-n\mathbbm{1}_{\{Q<0\}})\le\rho(0)=0$, and hence
$$\alpha_\rho(Q)\ge\E[-nQ\mathbbm{1}_{\{Q<0\}}]-\rho(-n\mathbbm{1}_{\{Q<0\}})\ge n\E[(-Q)_+],~~n\in\mathbb{N}.$$
Thus $\alpha_\rho(Q)=+\infty$. For any $z<0$, Assumption~\ref{ass:atomless} allows us to choose a two-point random variable $Q\in L^1_{\mathrm m}$ taking values $z$ and $b_0$. By Lemma~\ref{lem:oce-penalty-working},
$$
\infty=\alpha_\rho(Q)=\frac{b_0-1}{b_0-z}f(z)+\frac{1-z}{b_0-z}f(b_0).
$$
Since $f(b_0)<\infty$, this proves the claim. Scalar Fenchel--Moreau duality now gives
$$
\ell(x)=\sup_{z\ge0}\{xz-f(z)\},~~x\in\R.
$$
This is a supremum of affine functions with nonnegative slopes, so $\ell$ is nondecreasing. Therefore $\ell\in\mathcal L_+$.
\end{proof}

\begin{proof}[Proof of Proposition~\ref{prop:indep}]
We give four risk measures that each satisfy three of the axioms and violate the fourth.

The risk measure $\xi$ of Example~\ref{ex:premium} is an average of standard OCEs. Each satisfies Axioms~(SRA), (N), and (M) by Theorem~\ref{thm:oce}, and averaging preserves these properties. 
To see that $\xi$ violates Axiom~(FC), let $X=0$, let $Y$ be uniformly distributed on $\{0,1,2\}$, let $B$ be a fair coin that is independent of $Y$, and let $W_a=Ba+(1-B)(Y-a)$. Since $\mathrm{ES}_{1/2}(Y)=5/3$ and $\mathrm{ES}_{3/4}(Y)=2$, we have $\xi(Y)=11/6$. For every event $A$ with $\P(A)=1-p$, by the dual representation of Expected Shortfall, we have that  
$\mathrm{ES}_p(Z)\ge\E[Z\mathbbm{1}_A]/(1-p)$. With $A=\{B=0,Y=2\}\cup C$ for an event $C\subseteq\{B=1\}$ of probability $1/3$ and $1/12$, respectively, we obtain $\mathrm{ES}_{1/2}(W_a)\ge(2+a)/3$ and $\mathrm{ES}_{3/4}(W_a)\ge(4-a)/3$ for every $a\in\R$. Hence
$$\inf_{a\in\R}\xi(W_a)\ge1>\frac{11}{12}=\frac{\xi(X)+\xi(Y)}{2},$$
and $\xi$ violates Axiom~(FC).

Next, define
$$\chi(X)=\inf_{m\in\R}\left\{m+\E[(X-m)_+]+\P(X>m)\right\},~~X\in L^\infty.$$
This risk measure satisfies Axioms~(FC), (N), and (M) but violates Axiom~(SRA). To verify the first three properties, set $\ell(x)=x_++\mathbbm{1}_{\{x>0\}}$, so that $\chi(X)=\inf_{m\in\R}\{m+\E[\ell(X-m)]\}$. 
For every $X\in L^\infty$ and $m\in\R$, the inequality $\ell(x)\ge x$ gives $m+\E[\ell(X-m)]\ge\E[X]$. Taking the infimum over $m$ yields $\chi(X)\ge\E[X]$, while the choice $m=\operatorname{ess\,sup}X$ gives $\chi(X)\le\operatorname{ess\,sup}X$. Thus $\chi$ is real valued. Taking $X=c$ in these bounds gives $\chi(c)=c$ for every $c\in\R$, proving Axiom~(N).
Since $\ell$ is nondecreasing, $\chi$ satisfies Axiom~(M). The verification of Axiom~(FC) in the proof of Theorem~\ref{thm:main} uses only the infimum formula and not the convexity of the loss function, so $\chi$ satisfies Axiom~(FC).

To see that Axiom~(SRA) fails, let $E_2\subseteq E_1$ be events with $\P(E_1)=1/2$ and $\P(E_2)=1/4$, let $W=2\cdot\mathbbm{1}_{E_2}$, and let $\mathcal G=\sigma(E_1)$. Then $\E[W\mid\mathcal G]=\mathbbm{1}_{E_1}$. By conditional Jensen's inequality, $\E[W\mid\mathcal G]\lecx W$, so Axiom~(SRA) would imply $\chi(\E[W\mid\mathcal G])\le \chi(W)$. The choice $m=0$ gives
$$\chi(W)\le\E[W]+\P(W>0)=\frac12+\frac14=\frac34.$$
For the conditional mean, every cash amount $m\ge1$ gives $m+\E[\ell(\mathbbm{1}_{E_1}-m)]=m\ge1$. Every cash amount $m<1$ gives $\P(\mathbbm{1}_{E_1}>m)\ge\P(E_1)=1/2$, and since $m+\E[(\mathbbm{1}_{E_1}-m)_+]\ge\E[\mathbbm{1}_{E_1}]=1/2$,
$$m+\E[\ell(\mathbbm{1}_{E_1}-m)]=m+\E\left[(\mathbbm{1}_{E_1}-m)_+\right]+\P(\mathbbm{1}_{E_1}>m)\ge\frac12+\frac12=1.$$
Hence $\chi(\E[W\mid\mathcal G])\ge1>3/4\ge \chi(W)$.

Next, consider the functional $2\E$. If $X\lecx Y$, then $\E[X]=\E[Y]$, so $2\E$ satisfies Axiom~(SRA). For every transfer $a$, $\E[B(X+a)+(1-B)(Y-a)]=\frac12\E[X]+\frac12\E[Y]$, so $2\E$ satisfies Axiom~(FC). It clearly satisfies Axiom~(M). It violates Axiom~(N).

Finally, define $\rho_{\mathrm{MV}}(X)=\E[X]+\operatorname{Var}(X)$ for $X\in L^\infty$. Since
$$\inf_{m\in\R}\{m+\E[(X-m)+(X-m)^2]\}=\E[X]+\inf_{m\in\R}\E[(X-m)^2]=\rho_{\mathrm{MV}}(X),$$
this is the OCE with loss function $\ell(x)=x+x^2$, and it satisfies Axioms~(SRA), (FC), and (N) by Theorem~\ref{thm:main}. Let $X$ take the values $0$ and $4$ with equal probability. Then $X\le4$, but
$$\rho_{\mathrm{MV}}(X)=2+4=6>4=\rho_{\mathrm{MV}}(4),$$
so Axiom~(M) fails.
\end{proof}

\setcounter{equation}{0}
\setcounter{lemma}{0}
\setcounter{definition}{0}
\section{Proofs for Section~\ref{sec:selecting}}\label{app:B}

This appendix contains the proofs of the results of Section~\ref{sec:selecting}. We call a risk measure \emph{monetary} if it is monotone and cash additive.

\begin{lemma}\label{lem:mvrep}
Let $\rho: L^\infty \to \R$ satisfy Axioms~(SRA), (FC), and (N) and Property~(ER). Then there is a convex function $\psi:\R\to\R$ with $\psi(0)=0$ such that $\rho(X)=\E[\psi(X)]$ for every $X\in L^\infty$ with $\E[X]=0$.
\end{lemma}

\begin{proof}
By Proposition~\ref{prop:properties}, $\rho$ is law invariant and concave in probabilistic mixtures. Together with Property~(ER), this makes $\widehat\rho$ midpoint affine on distributions with mean zero. 
We claim that this midpoint affinity extends to affinity for every mixture weight in $[0,1]$. To see this, fix two distributions with mean zero $\mu,\nu$ and $t\in[0,1]$, and let $\mu_t=t\mu+(1-t)\nu$. Applying midpoint affinity twice gives
\begin{equation}\label{eq:mv-affinity-sum}
\widehat\rho(\mu_t)+\widehat\rho(\mu_{1-t})=2\widehat\rho((1/2)\mu+(1/2)\nu)=\widehat\rho(\mu)+\widehat\rho(\nu).
\end{equation}
Concavity in probabilistic mixtures gives
$$\widehat\rho(\mu_t)\ge t\widehat\rho(\mu)+(1-t)\widehat\rho(\nu).$$
Suppose this inequality were strict. Since concavity also gives $\widehat\rho(\mu_{1-t})\ge(1-t)\widehat\rho(\mu)+t\widehat\rho(\nu)$, adding the two inequalities would yield
$$\widehat\rho(\mu_t)+\widehat\rho(\mu_{1-t})>\widehat\rho(\mu)+\widehat\rho(\nu),$$
contradicting \eqref{eq:mv-affinity-sum}. Thus
$$\widehat\rho(t\mu+(1-t)\nu)=t\widehat\rho(\mu)+(1-t)\widehat\rho(\nu),~~t\in[0,1].$$

For $z<0$, let $\tau_z$ be the probability distribution with mean zero supported on $\{z,1\}$; for $z>0$, let it be the probability distribution with mean zero supported on $\{-1,z\}$; and set $\tau_0=\delta_0$. Write $w_z=\tau_z(\{z\})=1/(1+|z|)$ and $\tau_{-1}=\tau_1$, which we denote by $\tau$. Define $\psi(0)=0$ and
$$
\psi(z)=\begin{cases}
\widehat\rho(\tau_z)/w_z,&z<0,\\[2pt]
\bigl(\widehat\rho(\tau_z)-2(1-w_z)\widehat\rho(\tau)\bigr)/w_z,&z>0.
\end{cases}
$$
Then $\psi(-1)=2\widehat\rho(\tau)$ and $\psi(1)=0$, so
$$\int_{\R}\psi\,\mathrm d\tau_z=\widehat\rho(\tau_z),~~z\in\R.$$
For $z=0$, this identity uses Axiom~(N).

Let $\mu$ be a finitely supported distribution with mean zero, and write its support as $\{z_1,\ldots,z_K\}$. Set
$$\beta=\left(\sum_{k=1}^K\frac{\mu(\{z_k\})}{w_{z_k}}\right)^{-1}~~\text{and}~~\lambda_k=\frac{\beta\mu(\{z_k\})}{w_{z_k}},~~k=1,\ldots,K.$$
We have $\beta\in(0,1]$ and $\sum_{k=1}^K\lambda_k=1$. We claim that
$\sum_{k=1}^K\lambda_k\tau_{z_k}=\beta\mu+(1-\beta)\tau.$
To see this, use $\lambda_k w_{z_k}=\beta\mu(\{z_k\})$ to write
$$\pi=\sum_{k=1}^K\lambda_k\bigl(\tau_{z_k}-w_{z_k}\delta_{z_k}\bigr)=\sum_{k=1}^K\lambda_k\tau_{z_k}-\beta\mu.$$
For $z_k<0$, the measure $\tau_{z_k}-w_{z_k}\delta_{z_k}$ equals $(1-w_{z_k})\delta_1$; for $z_k>0$, it equals $(1-w_{z_k})\delta_{-1}$; and for $z_k=0$, it is zero. Hence $\pi$ is a nonnegative measure supported on $\{-1,1\}$. Since $\mu$ and all the $\tau_{z_k}$ have total mass $1$ and mean zero, $\pi=(1-\beta)\tau$, which proves the claim. By affinity and the identities for $\tau_z$,
$$
\begin{aligned}
\beta\widehat\rho(\mu)+(1-\beta)\int_{\R}\psi\,\mathrm d\tau
&=\beta\widehat\rho(\mu)+(1-\beta)\widehat\rho(\tau)\\
&=\sum_{k=1}^K\lambda_k\widehat\rho(\tau_{z_k})\\
&=\sum_{k=1}^K\lambda_k\int_{\R}\psi\,\mathrm d\tau_{z_k}=\beta\int_{\R}\psi\,\mathrm d\mu+(1-\beta)\int_{\R}\psi\,\mathrm d\tau.
\end{aligned}
$$
Therefore, $\widehat\rho(\mu)=\int_{\R}\psi\,\mathrm d\mu$. Thus $\rho(X)=\E[\psi(X)]$ for every simple $X$ with $\E[X]=0$.

To show that $\psi$ is convex, let $s<u<t$ with $u=\theta s+(1-\theta)t$ and $\theta\in(0,1)$. Choose a simple $X$ with mean zero and $\eta=\P(X=u)>0$; for instance, we can find a random variable with the distribution $\tau_u$. By Assumption~\ref{ass:atomless}, we can find $A,B\subseteq \{X=u\}$ with  
$A\cup B=\{X=u\}$, $A\cap B=\varnothing$, $\P(A)=\theta\eta$, and $\P(B)=(1-\theta)\eta$. Define a simple random variable $Z$ that equals $s$ on $A$, $t$ on $B$, and agrees with $X$ elsewhere. It is easy to see that $\mathbb{E}[Z]=0$ and that $\E[Z\mid X]=X$, so $X\lecx Z$ by conditional Jensen's inequality. Axiom~(SRA) and the representation just proved give $\E[\psi(X)]\le\E[\psi(Z)]$. Therefore
$$\mathbb{E}[\psi(X)\mathbbm{1}_{\{X\neq u\}}]+\eta\psi(u)=\mathbb{E}[\psi(X)]\le\mathbb{E}[\psi(Z)]=\mathbb{E}[\psi(X)\mathbbm{1}_{\{X\neq u\}}]+\eta\bigl(\theta\psi(s)+(1-\theta)\psi(t)\bigr).$$
Since $\eta>0$, this proves convexity.

Finally, let $X\in L^\infty$ with $\E[X]=0$, and choose a sequence $(Y_n)_{n\in\mathbb{N}}$ of simple random variables with $\lim_{n\to\infty}\|Y_n-X\|_\infty=0$. Define the sequence $(X_n)_{n\in\mathbb{N}}$ by $X_n=Y_n-\E[Y_n]$. Each $X_n$ is simple and $\E[X_n]=0$. Since $|\E[Y_n]|\le\E[|Y_n-X|]\le\|Y_n-X\|_\infty$, the triangle inequality gives
$$\|X_n-X\|_\infty=\|Y_n-X-\E[Y_n]\|_\infty\le\|Y_n-X\|_\infty+|\E[Y_n]|\le2\|Y_n-X\|_\infty.$$
It follows that $\lim_{n\to\infty}\|X_n-X\|_\infty=0$. 
This convergence implies that there is a constant $K>0$ such that $\|X\|_\infty\le K$ and $\|X_n\|_\infty\le K$ for every $n\in\mathbb{N}$. Since $\psi$ is finite and convex, it is uniformly continuous on $[-K, K]$. By the norm continuity established in the proof of Proposition~\ref{prop:properties}, we obtain
$$\rho(X)=\lim_{n\to\infty}\rho(X_n)=\lim_{n\to\infty}\E[\psi(X_n)]=\E[\psi(X)].$$
This establishes the representation for every $X\in L^\infty$ with $\E[X]=0$.
\end{proof}

\begin{proof}[Proof of Theorem~\ref{thm:mv}]
First, let $\rho(X)=\E[X]+\kappa\operatorname{Var}(X)$ for some $\kappa\ge0$. This is the OCE with loss function $\ell(x)=x+\kappa x^2$, so it satisfies Axioms~(SRA), (FC), and (N) by Theorem~\ref{thm:main}. 
If $\E[X]=\E[Y]=m$ and $B$ is a fair coin independent of $(X,Y)$, then $\E[BX+(1-B)Y]=m$. Moreover, $B(1-B)=0$ gives
$$\bigl(BX+(1-B)Y-m\bigr)^2=B(X-m)^2+(1-B)(Y-m)^2.$$
Taking expectations and using independence yields
$$\operatorname{Var}(BX+(1-B)Y)=\frac12\operatorname{Var}(X)+\frac12\operatorname{Var}(Y).$$
Thus $\rho(BX+(1-B)Y)=(1/2)\rho(X)+(1/2)\rho(Y)$, proving Property~(ER).

Conversely, let $\rho$ satisfy the three axioms and Property~(ER), and let $\psi$ be the convex function in Lemma~\ref{lem:mvrep}. Since $\psi$ is finite and convex on $\R$, \citet[Theorem~24.1]{Rockafellar1970} ensures that its left and right derivatives $\psi'_-$ and $\psi'_+$ exist and are finite everywhere.

We claim that the zero transfer minimizes
$$\Psi(a)=\E[\psi(X+a)]+\E[\psi(Y-a)]$$
whenever $X, Y$ are simple and have mean zero. To see this, choose a fair coin $B$ independent of $(X,Y)$. The loss $W_a=B(X+a)+(1-B)(Y-a)$ is simple with mean zero, and Lemma~\ref{lem:mvrep} gives
$$\rho(W_a)=\frac12\E[\psi(X+a)]+\frac12\E[\psi(Y-a)]=\frac12\Psi(a).$$
Axiom~(FC) therefore gives $\inf_{a\in\R}\Psi(a)=\rho(X)+\rho(Y)=\Psi(0)$, proving the claim.

Since $X,Y$ are simple, the right derivative of $\Psi$ at zero is $\E[\psi'_+(X)]-\E[\psi'_-(Y)]$. It is nonnegative because zero minimizes $\Psi$, so
\begin{equation}\label{eq:mvcomp}
\E[\psi'_-(Y)]\le\E[\psi'_+(X)]
\end{equation}
for all simple $X,Y\in L^\infty$ with mean zero. 
Define
$$c=\sup\bigl\{\E[\psi'_-(Y)]:Y\in L^\infty\text{ is simple and }\E[Y]=0\bigr\}.$$
Clearly, $c\ge\psi'_-(0)$. Taking $X=0$ in \eqref{eq:mvcomp} gives $c\le\psi'_+(0)$. Thus $c$ is finite.

Now fix a simple $Z$ with mean zero. The definition of $c$ gives $\E[\psi'_-(Z)]\le c$. On the other hand, setting $X=Z$ in \eqref{eq:mvcomp} and taking the supremum over $Y$ gives $c\le\E[\psi'_+(Z)]$. Hence
\begin{equation}\label{eq:mvsub}
\E[\psi'_-(Z)]\le c\le\E[\psi'_+(Z)]
\end{equation}
for every simple $Z$ with mean zero.

Let $\mathcal N$ be the set of points where $\psi$ is not differentiable. Since $\psi$ is finite and convex on $\R$, \citet[Theorem~25.3]{Rockafellar1970} implies that $\mathcal N$ is at most countable. 
For $x<0<y$ outside $\mathcal N$, choose $Z$ taking values $x$ and $y$ with probabilities $y/(y-x)$ and $-x/(y-x)$, respectively. Then $\E[Z]=0$, and \eqref{eq:mvsub} gives
$$\frac{y\psi'(x)-x\psi'(y)}{y-x}=c\implies\frac{\psi'(x)-c}{x}=\frac{\psi'(y)-c}{y}.$$
Fixing one negative and one positive differentiability point shows that these ratios equal a single constant, which we denote by $2\kappa$. Thus $\psi'(z)=c+2\kappa z$ for $z\notin\mathcal N\cup\{0\}$. 
Since $\psi$ is finite and convex on $\R$, by \citet[Corollary~24.2.1]{Rockafellar1970}, we have
$$\psi(z)=\int_0^z\psi'_+(u)\,\mathrm du=\int_0^z(c+2\kappa u)\,\mathrm du=cz+\kappa z^2,~~z\in\R.$$
Convexity implies $\kappa\ge0$.

For $X\in L^\infty$ with mean zero, Lemma~\ref{lem:mvrep} now gives $\rho(X)=c\E[X]+\kappa\E[X^2]=\kappa\operatorname{Var}(X)$. For an arbitrary $X\in L^\infty$, cash additivity from Proposition~\ref{prop:properties} gives
$$\rho(X)=\E[X]+\rho(X-\E[X])=\E[X]+\kappa\operatorname{Var}(X).$$
This proves the proposition.
\end{proof}

\begin{proof}[Proof of Proposition~\ref{prop:homogeneous}]
By Theorem~\ref{thm:oce}, $\rho=\rho_\ell$ for some $\ell\in\cL_+$. Assume that $\rho\ne\E$ and that $\ell$ is not finite-valued. For each $t>0$, define $\ell_t(x)=\ell(tx)/t$. Then $\ell_t\in\cL_+$, and positive homogeneity gives
$$
\rho_{\ell_t}(X)=\frac{\rho_\ell(tX)}{t}=\rho_\ell(X).
$$
By the uniqueness statement in Theorem~\ref{thm:main}, there is $c_t\in\R$ such that
$
\ell_t(x)=\ell(x+c_t)-c_t.
$
Since $\ell$ is nondecreasing and $\ell(0)=0$, we have that $
b=\sup\{x:\ell(x)<\infty\}\ge0.
$
Therefore, since $\ell_t(x)=\ell(tx)/t=\ell(x+c_t)-c_t$, we have that $b/t=b-c_t$ or $c_t=b-b/t$. 

For the sake of contradiction, assume that $b>0$. If $x\in (-\infty,b)$, we can find a $t>0$ such that $b-b/t=x$. Therefore, $$\ell(x)=\ell(b-b/t)=\ell(c_t)=c_t=x.$$
For $X\in L^\infty$, choose $m\in\mathbb{R}$ satisfying $X-m<b$. Then
$m+\E[\ell(X-m)]=\E[X].$
Since $\ell(x)\geq x$ for all $x\in \mathbb{R}$, this would imply $\rho=\E$, a contradiction.

Thus $b=0$, and $c_t=0$ for every $t>0$. Therefore, $\ell(x)=\ell(x+c_t)-c_t=\ell_t(x)=\ell(tx)/t$ for all $t>0$. It follows that
$$
\ell(x)=
\begin{cases}
ax,&x\le0,\\
+\infty,&x>0,
\end{cases}
~~ 0\le a<1.
$$
Where the fact that $\ell\in \mathcal{L}_+$ and $\ell(x)\geq x$ for all $x\in \mathbb{R}$ gives $a\in [0,1]$, and $\rho\neq \mathbb{E}$ gives $a\neq 1$.

Consequently,
$$
\rho(X)=\inf_{m\ge\operatorname{ess\,sup}X}
\{a\E[X]+(1-a)m\}
=a\E[X]+(1-a)\ES_1(X).
$$
This establishes the claimed family, including the extended-loss endpoint. The converse is trivial.
\end{proof}

\begin{proof}[Proof of Corollary~\ref{cor:ES}]
This follows from Proposition \ref{prop:homogeneous} and \citet[Theorem~3.1]{AmaranteLiebrich2024}.
\end{proof}

\begin{proof}[Proof of Proposition~\ref{prop:entropic}]
Suppose first that $\rho$ satisfies Axioms (SRA), (FC), (N), and (M) and
Property (FI). By Proposition~\ref{prop:properties} and Axioms~(N) and (M), $\rho$ is a normalized, law-invariant monetary risk measure that is concave in probabilistic mixtures. We verify the two remaining hypotheses of \citet[Theorem~6.1]{Embrechts2021}: betweenness and continuity at constant distributions.

If $\widehat\rho(\mu)=\widehat\rho(\nu)$, then Property~(FI), applied in both directions, gives
\begin{equation}\label{eq:fi-substitution}
\widehat\rho((1/2)\mu+(1/2)\eta)
=
\widehat\rho((1/2)\nu+(1/2)\eta),
\qquad \eta\in\mathcal D^\infty.
\end{equation}
Taking $\eta=\nu$ shows that the level set containing $\mu$ and $\nu$ is midpoint convex. By induction,
$$
\widehat\rho(t\mu+(1-t)\nu)=\widehat\rho(\mu)
$$
for every dyadic $t\in[0,1]$. Since $\widehat\rho$ is finite and concave in probabilistic mixtures, it is continuous along this mixture line on $(0,1)$. The equality therefore holds for every $t\in[0,1]$. Thus $\rho$ satisfies betweenness.

For continuity, fix $a>0$. By Assumption~\ref{ass:atomless}, choose a fair coin $B$ and an independent uniform random variable $U$ on $[0,1]$, and set $B_t=\mathbbm{1}_{\{U\le t\}}$. Monotonicity and $0\le aB_t\le a$ imply that $\lim_{t\downarrow0}\rho(aB_t)=r$ exists in $[0,a]$. We show that $r$ cannot be positive.

Since $aBB_t$ has the same law as $aB_{t/2}$, Property~(FI) allows us to replace the branch $aB_t$ by the sure loss $\rho(aB_t)$ in a fair-coin mixture with common branch zero. Hence
$$
\rho(aB_{t/2})=\rho(aBB_t)=\rho(B\rho(aB_t)).
$$
Halving $t$ does not change the limit on the left. On the right, the scalar $\rho(aB_t)$ converges to $r$, so $B\rho(aB_t)$ converges to $rB$ in $L^\infty$. By the norm continuity established in the proof of Proposition~\ref{prop:properties}, taking limits gives $r=\rho(rB)$. If $r>0$, Property~(FI) applied to the constants $0<r$ with common third loss $r$ gives
$$
\rho(rB)=\rho((1-B)r)<\rho(r)=r,
$$
a contradiction. Thus $r=0$, and cash additivity gives
\begin{equation}\label{eq:fi-continuity}
\lim_{t\downarrow0}\widehat\rho((1-t)\delta_x+t\delta_y)
=
x+\lim_{t\downarrow0}\rho((y-x)B_t)
=x,
\end{equation}
for every $x<y$. This is the continuity at constant distributions of the cited theorem. Since Property~(FI) yields betweenness and the continuity condition also holds, \citet[Theorem~6.1]{Embrechts2021} gives $\rho=\rho_\gamma$ for some $\gamma\ge0$.

Conversely, let $\rho=\rho_\gamma$ for some $\gamma\ge0$. For $\gamma>0$, this is the standard OCE with loss function $\ell(x)=(e^{\gamma x}-1)/\gamma$; for $\gamma=0$, use $\ell(x)=x$. Thus $\rho$ satisfies the four axioms by Theorem~\ref{thm:oce}. To verify Property~(FI), let $B$ be a fair coin independent of $(X,Y,Z)$. For $\gamma>0$, the definition and independence give
$$
\rho_\gamma(BX+(1-B)Z)
=
\frac1\gamma\log\left(
\frac12e^{\gamma\rho_\gamma(X)}
+\frac12e^{\gamma\rho_\gamma(Z)}
\right).
$$
For fixed $Z$, the right-hand side is strictly increasing in $\rho_\gamma(X)$, so comparing the mixtures with $X$ and $Y$ is equivalent to comparing $\rho_\gamma(X)$ and $\rho_\gamma(Y)$. For $\gamma=0$, the same equivalence follows from
$$
\rho_0(BX+(1-B)Z)=\frac12\E[X]+\frac12\E[Z].
$$
This proves the proposition.
\end{proof}
\end{document}